%% file: Amp.tex
\RequirePackage{color}
\documentclass{IEEEtran}

\usepackage{comment}
\usepackage{xspace}
\usepackage{fancyvrb}
\usepackage{verbatimbox}
\usepackage{graphicx}
\usepackage{url}
\usepackage{xspace}
\usepackage[normalem]{ulem}
\usepackage{marginnote}
\usepackage{comment}
\usepackage{multirow}
\usepackage{tabularx}
\usepackage{booktabs}
\usepackage{cleveref}
\AtBeginDocument{%
  \providecommand\BibTeX{{%
    \normalfont B\kern-0.5em{\scshape i\kern-0.25em b}\kern-0.8em\TeX}}}

\def\Sys{AMP\xspace}

\def\SysServ{\textit{\Sys Service}\xspace}
\def\ManDb{\textit{Manifest Database}\xspace}
\def\CcfLed{\textit{Media Provenance Ledger}\xspace}
\def\Tools{\textit{\Sys Authoring Tools}\xspace}

\def\Man{manifest\xspace}
\def\Mans{manifests\xspace}

\begin{document}

\title{\Sys: Authentication of Media via Provenance}
%\author{Anonymous Submission to MMSys20}
\author{\IEEEauthorblockN{Paul England, Henrique S. Malvar, Eric Horvitz, Jack W. Stokes, C\'edric Fournet, Rebecca Burke-Aguero, \\
Amaury Chamayou, Sylvan Clebsch, Manuel Costa, John Deutscher, Shabnam Erfani, Matt Gaylor, \\
Andrew Jenks, Kevin Kane, Elissa Redmiles, Alex Shamis, Isha Sharma, Sam Wenker, Anika Zaman}

\IEEEauthorblockA{\textit{Microsoft}}}
%\and
%\and
%\and
%\and
%\and
\maketitle

\input{abstract}

\input{intro}
\input{system}

\input{manifest}
\input{binding}
\input{ccf}
\input{watermark}

\input{service}

\input{uxSupp}
\input{tools}
\input{flow}
\input{eval}

\input{consortium}
\input{discussions}
\input{related}
\input{conc}

\bibliographystyle{IEEEtran}
\bibliography{Amp}

\appendices
\input{introSupp}
\input{manifestSupp}
\input{mandet}

\input{ccfSupp}

\end{document}

%% file: abstract.tex
\begin{abstract}
Advances in graphics and machine learning have led to the general availability of easy-to-use tools for modifying and synthesizing media. The proliferation of these tools threatens to cast doubt on the veracity of all media. One approach to thwarting the flow of fake media is to detect modified or synthesized media through machine learning methods. While detection may help in the short term, we believe that it is destined to fail as the quality of fake media generation continues to improve. Soon, neither humans nor algorithms will be able to reliably distinguish fake versus real content. Thus, pipelines for assuring the source and integrity of media will be required---and increasingly relied upon. We propose \Sys, a system that ensures the authentication of media via certifying provenance. \Sys creates one or more publisher-signed manifests for a media instance uploaded by a content provider. These manifests are stored in a database allowing fast lookup from applications such as browsers. For reference, the manifests are also registered and signed by a permissioned ledger, implemented using the Confidential Consortium Framework (CCF). CCF employs both software and hardware techniques to ensure the integrity and transparency of all registered manifests. \Sys, through its use of CCF, enables a consortium of media providers to govern the service while making all its operations auditable. The authenticity of the media can be communicated to the user via visual elements in the browser, indicating that an \Sys manifest has been successfully located and verified.
\end{abstract}

%% file: intro.tex
\section{Introduction}
\label{sec:intro}
Advances in graphics and machine learning have enabled the creation and distribution of easy-to-use tools for synthesizing fake media. These tools enable non-expert users to modify or synthesize audiovisual media that looks convincingly real. Although subtle artifacts may be detected in some cases by experts or by statistical classifiers developed with machine learning, we expect that the march of technical advances will soon make it impossible to distinguish fake media from real. Tools for media synthesis, coupled with wide-scale distribution of social media, threaten to cause harm to individuals, institutions, and nations. More generally, widespread distribution of fake media has the potential to undermine society’s trust in the veracity of all media. With the rise of fake media, what can be done to protect the veracity of media and provide a pathway to trust?

We are pursuing an answer by providing users with reliable information about the source and authenticity of a media object, through a verifiable and trustworthy media authentication service. That should allow the consumer to rely on the reputation of the media producer to make informed decisions about the media’s trustworthiness. For example, a media company or publisher can attest that it published a work in accordance with their editorial standards, or content captured at a certain location and time by cameras in the hands of a trusted reporting team.

The simplest building block for proving provenance is to sign the media object digitally. However, the variety of mechanisms for media distribution, with many of them modifying the media files or streams, means that maintaining digital signatures is difficult. Additional challenges are also involved. For example, in a typical redistribution scenario, media content is re-encoded by a content distribution network (CDN). Such re-encodings are needed to address variations in channel bandwidth, rendering device resolution, and other constraints. To preserve provenance information, certificates must be tracked and re-inserted for each transformation.

We present a practical system named \Sys (for \textit{a}uthentication of \textit{m}edia via \textit{p}rovenance) aimed at providing robust verification of provenance while supporting a wide variety of production and distribution scenarios at Internet scale. We propose \Sys as an approach to mitigating the negative societal impact of fake/synthetic media, based on certifiable provenance. The \Sys effort brings together expertise in security and media, leveraging advances in cryptography, watermarking, and recently released cloud security and ledger services.

Threats to the integrity of sources include the use of a range of techniques, from simple modifications of timing to more sophisticated uses of graphics and generative models, for manipulating or synthesizing audiovisual content that is perceived by consumers as capturing actual events.

Approaches to securing media from a reputable provider to its consumption include (1) strong authentication and (2) fragile watermarking.  A complementary approach involves (3) the detection of manipulation or synthesis via pattern recognition employing machine-learned classifiers.
Additional opportunities include (4) event-certification methods for certifying that media as captured is linked to actual physical events, rooted in activities that are certified via a combination of methods to have occurred at a time and place. With \Sys, we focus on securing media based on the joint use of (1) and (2), thereby providing the certification of the identity of the media provider.

The \Sys system consists of four main modules including
the \SysServ,
the \CcfLed,
the \ManDb, and
the \Tools.
\Sys authenticates media using a digitally signed data structure called a \emph{manifest}, and the \SysServ allows content providers to upload their media manifests to \Sys.  Manifests are registered in the \emph{Media Provenance Ledger}, which is a public distributed ledger based on the Confidential Consortium Framework (CCF)~\cite{CcfTech,CcfDoc}. Manifests can be distributed together with media contents, whereas the ledger ensures integrity and auditability of the full history of media publishing operations.
In addition, manifests are indexed by media fragments in a \emph{Manifest Database} for fast querying.  Once a manifest or group of manifests has been uploaded, media players can then use the \SysServ to validate the authenticity of the corresponding media contents, even if the content is distributed without its manifest.
A set of tools allows content providers to interact with the \SysServ when the content is published.  In addition to the service and tools, media players (browsers, smartphone applications, etc.) need to be extended to check and display provenance information.

Enabling large-scale media provenance will require the cooperation of multiple participants, including content producers, publishers, and technology providers. We envision \Sys supporting a media provenance consortium with open governance rules, where all the governance operations are recorded in the \CcfLed, for auditability and transparency. We hope that the \Sys project will be a starting point for broadly adopted standards for media provenance verification.

In this paper we make the following novel contributions:
\begin{enumerate}
  \item We describe an end-to-end solution for media producers to provide provenance information for each media item produced.
  \item We describe and benchmark a novel system to track the provenance of videos uploaded to the Internet.
  \item We describe how these videos can be distributed via CDNs or social media platforms while maintaining the required provenance information and not requiring coordination with the CDN providers.
  \item We show how the use of novel ledger techniques can scale-out to handle the majority of media items produced for distribution on the Internet.
\end{enumerate}

%% file: system.tex
\section{\Sys System Overview}
\label{sec:system}
\begin{figure}[tb]
\centering
  \includegraphics[trim = 1.0in 3.0in 0.2in 3.0in,clip,width=1.0\columnwidth]{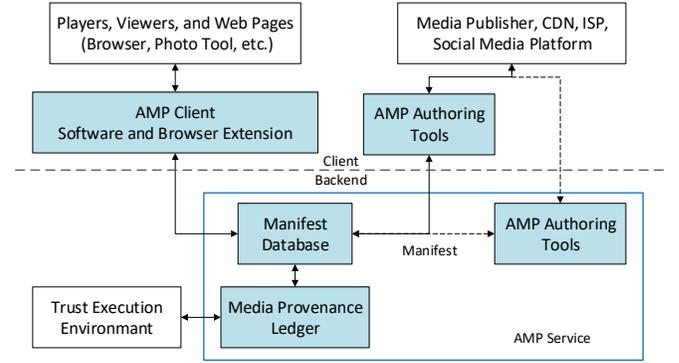}
  \caption{High-level overview of the \Sys system components (shown in blue).}
  \label{fig:system}
\end{figure}

We provide an overview in this section of the core \Sys concepts and how they are composed to form an end-to-end media authentication and verification system.
Figure~\ref{fig:system} illustrates how the \Sys components are integrated into a production, distribution, and rendering pipeline. A content provider uses the \Tools to create signed manifests and register them as part of publication, so that it can be authenticated by the \SysServ. The manifests can either be created locally by the publisher if they do not want to upload the media to the backend or alternatively in the \SysServ itself. Other organizations such as a CDN, social media platform, or internet service provider (ISP) can similarly record transformations that they apply to the content provider's original media content, using the \Tools.
The \SysServ records the resulting publication metadata in a manifest, signed by the provider (or the transformer), and stored in a \ManDb (DB) for fast verification. One or more cryptographic hashes of the media content are also stored in a verifiable ledger, called the \CcfLed, using CCF. Finally, consumer applications such as browsers, web sites, or media players use \Sys manifests and libraries for verifying (i.e., authenticating) that a media item indicated as coming from a content provider has been previously registered in the \SysServ by that provider.

\subsection{\Sys System Components}

\noindent\textbf{\Sys Manifest.}
The manifest is the central data structure in \Sys. It authenticates media objects (including various cryptographic hashes of their encodings) and binds them to their publisher-provided metadata.
Manifests support simple media objects, streaming media, progressive download
%and audio,
and adaptive bitrate streaming. A manifest can also record the attribution of derived works through ``back-pointers'' to one or more source objects, as well as descriptions of how the original works were transformed.

\Sys includes two different sets of modules, one for a near-term solution and another for a long-term solution. The near-term modules indicate provenance using a detached manifest and can work with today's media formats and infrastructure. In addition, the long-term solution uses an embedded manifest, which is included in the media stream itself, but it will ultimately require extensions to both media and browser standards.

\noindent\textbf{Media Provenance Ledger.}
Manifests are recorded on a CCF blockchain. CCF operates the ledger (i.e., blockchain) of published works, which is essentially a list of manifests, relying on trusted hardware and providing high availability via the Raft~\cite{RAFT} consensus protocol.
Our implementation of CCF supports the registration of new manifests and issues signed manifest receipts. These receipts complement the producer's signatures; they enable any media consumers to independently verify that the media they receive  has been published with the corresponding metadata. CCF natively supports online querying and validation of transactions along with their endorsing certificates.

\noindent\textbf{Manifest Database.}
Eventually in the long-term solution, we expect manifests to be distributed with the media objects themselves so that their provenance can be verified locally. To support a gradual transition, and to withstand the distribution of media without their associated manifests (e.g., media streamed from YouTube), \Sys maintains an indexed manifest database, so that clients can retrieve manifests given media excerpts.

\noindent\textbf{\Sys Service.}
The \SysServ exposes the \ManDb and the \CcfLed to client application through a set of REST APIs.

\noindent\textbf{\Sys Tools and Libraries.}
We also provide a set of tools and libraries for interacting with the \SysServ. The tools cover: (a) the creation, signing, and ingestion of content/manifests into the \Sys system,  (b) querying the \Sys system for media authentication information and checking that media objects are intact,  and (c) \Sys service governance (adding/removing members and users, etc.).

\noindent\textbf{Fragile Watermarking.}
In many cases, the media will be transformed without registering a manifest that records the transformation.
To facilitate the retrieval of any manifest for the original media object, the publisher can insert a watermark using the \Sys Watermark Tool.
This watermark carries a unique manifest identifier, which may be used to retrieve the original contents and metadata, and can be used to compare them with the transformed media.

\noindent\textbf{User Experience Components.} To provide a good user experience, \Sys includes three components including two variants of a browser extension,
two demonstration web pages, and a modified Chromium browser capable of displaying a new variant of an HTML video element.

\input{implementation}

%% file: implementation.tex
\noindent\textbf{Implementation}
\label{sec:implementation}
\Sys has been implemented to run on Windows and Linux (Ubuntu LTS 18.04). The Media Provenance Ledger has been developed and tested on Ubuntu 18.04 since the CCF framework only currently supports this version of Linux.

The core \Sys components are primarily implemented in C\# using .NET Core 3.1 so that the system will
run on Linux, MacOS and Windows. The browser extension is implemented in JavaScript and HTML. CCF is primarily written in C++ although it allows applications to be written in either C++ or Lua.
In addition, the audio watermarking code is implemented in C. This implementation enables efficient porting to many different processing environments.

%% file: manifest.tex
\section{\Sys Manifests}
\label{sec:manifest}
%
%\Sys manifests are data structures that describe media objects.
An \Sys manifest is a data structure that cryptographically authenticates media objects and their associated metadata. Manifests
are registered on the Media Provenance Ledger (Section~\ref{sec:ccf}), optionally distributed by media providers and distributors,
and recorded in a complementary Manifest Database (Section~\ref{sec:db}).
% indexed for fast queries.
%in several ways.
The purpose of manifests is to allow media player clients to quickly and easily verify the publisher (and possibly the distributor) of a media object.
The values stored in the manifest data structure are generated by the content provider as it publishes
%and authenticate
the media object.

\Sys supports two types of manifests: static and streaming.
A static manifest handles a simple media object (e.g. JPEG) or a collection of objects with different encodings (facsimiles),
while a streaming manifest contains an array of cryptographic hashes corresponding to ``chunks'' of the associated media.
For example, a chunk might correspond to one or more seconds of video or audio.
%\Sys supports two types of manifests: static and streaming.
%A static manifest contains the cryptographic hash of its associated media object (e.g. a JPEG) or a collection of objects with different encodings (facsimiles).  A streaming manifest contains an array of cryptographic hashes corresponding to ``chunks'' of the associated media.  For example, a chunk might correspond to a few seconds of video or audio.

\Sys manifests can be used to authenticate the original source material, or the transformation from one format to another.
Note that checking whether a transformation is faithful is not discussed here.

\Sys manifests are signed by publishers, CDNs, etc.
%, and the cryptographic hash of the manifests are recorded on the ledger.
The cryptographic hash of a manifest is called its \Sys manifest ID (ManifestID).  It serves as a unique identifier and a commitment for the manifest. ManifestIDs are also digitally signed by content producers or distributors, and recorded on the ledger. \Sys uses X.509 to create all digital signatures and SHA-256 for all cryptographic hashes.

\begin{comment}
\subsection*{Authentication Challenges for Streaming Media}
\Sys authenticates media objects with digital signatures.  It is straightforward to do this with text and images: we simply generate the cryptographic hash and then sign picture.jpg or doc.html.  Streaming media is more problematic because (a) an application should not have to wait to download the entire file before it can check the signature, (b) streaming services support changing the stream resolution to match network constraints (adaptive bitrate streaming), (c) some transport layers are lossy, and (d) users can seek back and forth within streams.  These issues imply that \Sys must authenticate much smaller regions (i.e., ``chunks'') in the stream.
\end{comment}

%\subsection{Static Manifests}
\noindent\textbf{Static Manifests.}
\begin{comment}
An \Sys manifest is a data structure that cryptographically authenticates media objects and associated metadata.  There are two types of manifests: static and streaming.  A static manifest contains the cryptographic hash of its associated media object (e.g. a JPEG) or a collection of objects with different encodings (facsimiles).  A streaming manifest contains an array of cryptographic hashes corresponding to ``chunks'' of the associated media.  For example, a chunk might correspond to a few seconds of video or audio.

Manifests can be used to authenticate the original source material or can authenticate transformations from one format to another.  (Note that checking whether a transformation is faithful is not discussed here.)
\end{comment}
%
%The details of a static manifest are provided in Table~\ref{tab:static_man}.
A few of the important fields of a static manifest (and the streaming manifest) are provided in Table~\ref{tab:static_man}.
A detailed description of the static manifest can be found in the appendix.
%supplementary materials.
The publisher assigns a MediaID to identify a particular media object.
In addition, the MediaID is encoded into the media object as a watermark and may also be inserted into the media's metadata.

The EncodingInformation field contains a string which indicates the media type (e.g., ``JPEG'', ``MP4''). This field helps to guard against the media's cryptographic hashes being wrongly interpreted.

\Sys manifests can also authenticate media objects that are derived from other media objects by means of  ``back pointers'' to one or more source manifests.  These ``transformation manifests'' can be used by publishers or CDNs to record transcoding and re-compressions of source material.  Transformation manifests can also be used to record the original media objects that were edited together to make a composite derived work.

The value of the OriginManifestID field includes one or more ManifestIDs that describe the source media used to create a derived work.
If a media object is a simple transcoding of another media object, this will be a single element array.  If a media object is created from several source objects (e.g., a news video
created from several original media objects) then additional ManifestIDs can be recorded in the array.
Note that OriginManifestID[] is not authoritative on its own:  it should only be trusted if the ManifestID that describes the transform is signed by a trusted authority.

The \Sys manifest includes a Copyright field which can be used to provide the copyright string associated with
the media object. This field provides a simple and legally enforceable way of limiting fake or misleading manifests.
Allowed strings may also be dictated in the \Sys terms of service.

In the simplest case (e.g., a picture or a text file), the manifest contains the cryptographic hash of the image or text and its associated metadata in the ObjectHash array field.  Optionally, the publisher can create and authenticate more than one encoding of a media object to optimize for client screen resolutions or network conditions.  We call these alternate representations \emph{facsimiles}.

\begin{table*}
  \begin{center}
      \begin{tabular}{|l|l|l|}
        \hline
        Field & Manifest Type & Description \\
        % & Type &  \\
        \hline
        \hline
        MediaID & Static/Streaming & Publisher-assigned identifier for the media object.  \\
        \hline
        MasterCopyLocator & Static/Streaming & URI of a stable, publisher provided location service or a generic URL redirector service.\\
        \hline
        EncodingInformation & Static/Streaming & String describing the media type (e.g., ``JPEG'', ``MP4''). \\
        \hline
        OriginManifests[] & Static/Streaming &  One or more ManifestIDs that describe the source media used to create a derived work.  \\
        \hline
        %If a media object is a simple transcoding of another media object, this will be a single element array.  If a media object is created from several source objects (e.g., a news video
        %created from several original media objects) then additional ManifestIDs can be recorded in an array.
%Note that OriginManifests[] is not authoritative on its own:  It should only be trusted if the ManifestID that describes the transform is signed by an a trusted authority.
        Copyright & Static/Streaming & Copyright string associated with the media object.\\
        \hline
        ObjectHash[] & Static & Cryptographic hash of the associated simple media object (or collection of related media \\
        & & objects). \\
        ChunkDigest & Streaming & An ordered array of chunk-hashes starting from the beginning of the work. \\
        %All ChunkAuthenticators have a list of chunk digests, but specific authenticators may have additional data that describe exactly what each chunk maps to (e.g. file offset-based, I-frame-to-I-frame, etc.).\\
        \hline
      \end{tabular}
  \end{center}
  \caption{Key manifest fields. A more detailed description of the manifest structures can be found in the supplementary material.}
  \label{tab:static_man}
\end{table*}

\noindent\textbf{Streaming Manifests.}
\Sys authenticates media objects with digital signatures.  It is straightforward to do this with text and images: we simply generate the cryptographic hash and then sign picture.jpg or doc.html.  Streaming media is more problematic because (a) an application should not have to wait to download the entire file before it can check the signature,
(b) streaming services support changing the stream resolution to match network constraints (adaptive bitrate streaming), (c) some transport layers are lossy, and (d) users can often navigate back and forth in streams.  These issues imply that \Sys must authenticate much smaller regions (i.e., ``chunks'') in the stream.

All of the fields for the streaming manifest match those in the static manifest in Table~\ref{tab:static_man} with the exception of the final field. 
While a static manifest contains one or more cryptographic hashes of an image or text document in the ObjectHash field, a streaming manifest contains a ChunkDigest which includes an ordered array of chunk-hashes.
However, the details of the streaming manifest in the appendix should be consulted for more details.

Clients must be able to quickly determine where individual chunks start and end in order to be able to calculate the cryptographic hashes of the chunks and compare these against the entries in an \Sys manifest.  Unfortunately, different media formats and network delivery mechanisms require different chunking strategies.

In one case, the \Sys system supports file offset-based chunking, which works well for HTTP GET-based streaming (which is most common on today's internet).
Lossy broadcast streaming requires different chunking strategies, such as I-frame-to-I-frame chunks for an MPEG stream.
Practically, streaming players process a cryptographic hash of a chunk every few seconds.  In most scenarios, consecutive chunks delivered to the client will map to consecutive chunk-hashes in a single manifest.  However, if a server is dynamically switching streams, then more than one manifest may be needed to authenticate a stream.

\Sys also supports adaptive bitrate streaming protocols such as DASH and HLS.  Adaptive bitrate streaming requires several different encodings of a media object, optimized for different network conditions and client capabilities.  Adaptive bitrate streams are supported in \Sys either by publishing several  manifests authenticating the different encodings, or by using a single manifest that authenticates multiple facsimiles.

\noindent\textbf{Detached and Embedded Manifests.} Initially before encoding standards can be modified, manifests will be stored separately from the media itself, and we call these ``detached manifests''. Long-term, we hope
that ``embedded manifests'' will be contained within the media's metadata and be transported within the media stream itself. We have implemented two versions of \Sys utilizing both detached manifests and embedded manifests. 

%% file: binding.tex
\section{Provenance Binding}
\label{binding}
\begin{figure}[tb]
  \centering
  \includegraphics[trim = 0.1in 0.1in 0.1in 0.1in, clip,width=1.0\columnwidth]{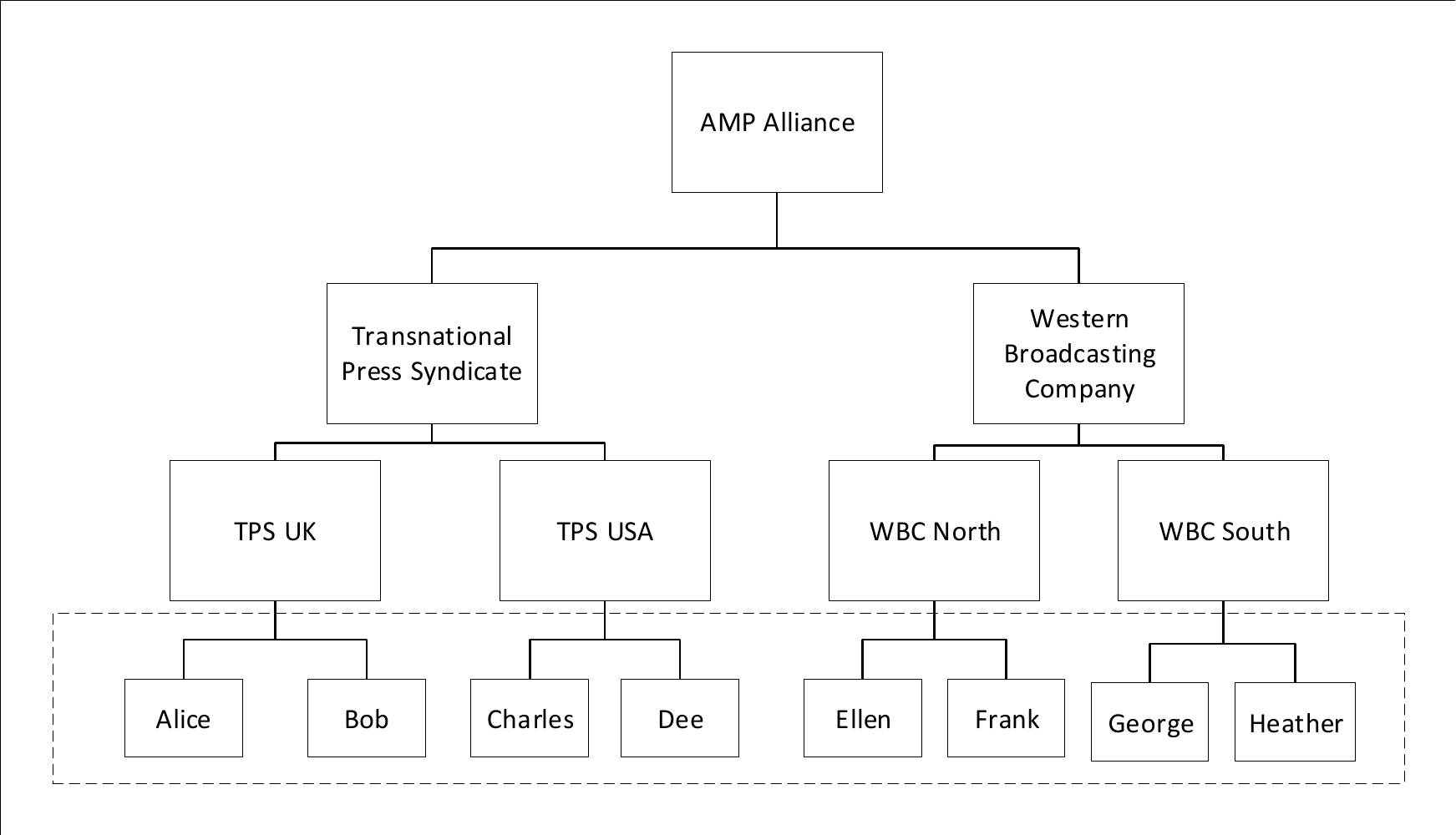}
  \caption{Example Public Key Infrastructure}
  \label{fig:pki}
\end{figure}
Authenticating that media has not been altered since the manifest was signed demonstrates the media's integrity, but tying the signer to an identity known and trusted by the consumer is what provides \emph{provenance}, and allows the consumer's trust in that producer to be imputed to the media. We have deployed a public key infrastructure (PKI) of X.509 certificates~\cite{x509} governed and administered by
an
alliance to provide a \emph{root of trust} for establishing identity. The alliance is then trusted to verify the identity of media producing organizations and individuals, and issue credentials from its Certificate Authority (CA) to those organizations and individuals that can be used to sign manifests and authenticate to the \SysServ. We expect that this responsibility will be delegated to Certificate Authorities, who already provide these services for the authentication of secure web sites. They will perform due diligence in establishing the identity of media producer applicants for credentials under contractual obligations to the alliance. The hierarchical nature of a certificate-based identity system allows a single parent credential to be issued to the organization, which can then issue subordinate credentials for individuals or organizational units. The exact structure of the subtree of the PKI for a particular organization is beyond the scope of this design, as it is intended to be customized to the particular needs and structure of each media producer.

Initially the root(s) of this PKI will be operated by the alliance and disconnected from the roots of trust currently used for the web PKI. Certificates used by participants will be given Extended Key Usage (EKU) extensions authorizing them for particular purposes. We have identified five uses and therefore five EKUs to use in this PKI: 1) server authentication, used by the \SysServ to authenticate itself to clients, 2) client authentication, used by clients to authenticate themselves to the \SysServ, 3) manifest signing, which will be used by producers to sign manifests, 4) time stamping, which will be used by the \SysServ and ledgers to attest to the publication time of a manifest, and 5) ledger registration, which will be used by ledgers to countersign manifests and attest they have been registered on that manifest. Server authentication, client authentication, and time stamping already have standard EKUs defined by the standard, and we will use those. Manifest signing and ledger registration EKUs are new purposes for which permanent, unique EKUs have not yet been allocated. We expect some certificates will be issued with multiple purposes: for example, the signer of a manifest will frequently be the client who registers it with the \SysServ, and so may use the same certificate for both purposes. Whether or not to combine these purposes in a single certificate becomes a governance decision for the alliance and media-producing entities, and the structure of our PKI allows for both possibilities.

One possible structure for such a PKI is given in Figure~\ref{fig:pki}. A single root operated by the alliance sits at the top, and issues intermediate CA credentials to each participating organization: in this example, the Transnational Press Syndicate (TPS) and the Western Broadcasting Company (WBC). These organizations each in turn issue further credentials to units of their organization: UK and USA bureaus in the case of TPS, and North and South bureaus in the case of WBC. Below each of these intermediates are individuals, but they are enclosed in a dotted-line box because, as described above, these are optional: An organization may wish to issue signing credentials to individuals, in which case the organizational unit credentials are also intermediate CAs. Alternatively, organizations may wish to maintain centralized publication pipelines, ingest media from individuals through a mechanism external to AMP, and sign as part of this process. In this case, the organizational unit credentials themselves are leaf certificates.

%% file: ccf.tex
\section{Media Provenance Ledger}
\label{sec:ccf}

\Sys{} implements an instance of CCF~\cite{CcfCode} to build a ledger-based application which is designed to securely store a cryptographic hash and copyright string for each manifest.
Any application built with CCF is designed to be administered by a group of consortium members via CCF's governance features.
Additionally, \Sys{} utilizes signed receipts as standalone proof that manifests are registered at a given index in the ledger.

CCF exposes to its users a key-value store.
This key-value store provides a simple abstraction of keys being a cryptographic hash of a manifest (i.e., ManifestID), with
the value being a signature computed by the publisher over a concatenation of the ManifestID and the copyright string (i.e., Copyright in Table~\ref{tab:static_man}).
Once written, these key-value pairs are stored in a Merkle tree, and the Merkle tree is replicated and stored on persistent storage.
To ensure that any tampering can be detected, CCF maintains a private key that the service protects and occasionally uses it to sign the Merkle root in the distributed ledger.

One of the core features that \Sys{} utilizes from CCF is its universally verifiable receipts.
The receipt for a given request validates the query, its response, and, more importantly, it certifies that its execution was recorded on the ledger.
The key proposition of a receipt is that it is possible to cryptographically validate that the signature of the manifest's cryptographic hash and the copyright string were successfully recorded, based on just the manifest, the receipt, and the public key of the CCF service~\cite{CcfTech} without needing to contact the CCF service.

Our prototype is designed to be run in a cloud datacenter.
In a real-world implementation we expect and have designed the service to be run by an operator (such as Azure).
CCF's utilization of trusted execution environments allows for a CCF service to be run in a public cloud while maintaining confidentiality from the cloud provider or operator.

%% file: watermark.tex
\section{Fragile Watermarking}
\label{sec:watermark}

We use watermarking to modify the media content in an imperceptible way. Faint noise-like patterns are inserted within the media content at production, and they can be read back at rendering. We tune the watermarking parameters such data media editing that preserves reasonably high fidelity preserves the detectability of watermarks, whereas heavier editing such as partial content replacement or fake media insertions~\cite{Deepfakes,FaceSwap}) will render the watermarking indetectable. Hence the term fragile watermarking.

We propose the use of fragile watermarking techniques using a spread-spectrum approach~\cite{malvar2003improved}, which adds low-level pseudorandom noise patterns within the media payload, be it video, audio, or images. The added noise is low enough (comparable to the small distortions due to the compression formats) and can be embedded in such a way that makes it imperceptible to human eyes and ears.

For each type of media and application scenario, we can design watermarking parameters that influence the thresholds on allowed changes, so that various kinds of minor modifications are considered as benign editing. In addition, we use keyless watermarking for \Sys which simplifies system design and makes watermarking detection open, so it can be performed by any entity in the media distribution path.

\noindent\textbf{Watermark Payload and Insertion.} The watermark payload string, which is inserted into the media item, is described in Table~\ref{tab:wm_payload} and contains the following fields: a media object ID (MediaID), a publisher URI (MasterCopyLocator), and a signature over these two fields (WatermarkPayloadSignature). \Sys does not provide a centralized database containing the MasterCopyLocator and MediaID. Instead after decoding, the client extracts the payload and submits the MediaID to the publisher using via MasterCopyLocator. Both the MediaID and the MasterCopyLocator are specified by the publisher. The MasterCopyLocator is typically a URI for the publisher’s Web service which is used to locate the media by their unique MediaID. The watermarking insertion process transforms a media object by embedding a signed watermark before its publication.

\noindent\textbf{Watermark Decoding.} The client inputs a media object to the Watermark Verification Module in the \Sys libraries to extract the watermark payload fields depicted in Table~\ref{tab:wm_payload}. The Watermark Verification Module uses the MasterCopyLocator to obtain a signing certificate. Then, the Watermark Verification Module uses this signing certificate to check the WatermarkPayloadSignature over the MediaID and MasterCopyLocator. If this cryptographic step succeeds, it finally returns the MediaID and the MasterCopyLocator back to the client. Once the client has recovered the MasterCopyLocator and the MediaID, it can then contact the publisher’s provenance service to authenticate that the media is valid. Watermark extraction is keyless: either it fails, or it returns the watermark payload.

\begin{table}
    \begin{center}
      \begin{tabular}{|l|l|}
        \hline
        Field & Description \\
        \hline
        \hline
        MediaID &  Publisher-assigned identifier;   \\
                 &  same as in Table~\ref{tab:static_man}. \\
        \hline
        MasterCopyLocator & Same as in Table~\ref{tab:static_man}. \\
        \hline
        Watermark & Signature value over the MediaID \\
        Payload   & and the PublisherID.  \\
        Signature &   \\
        \hline
      \end{tabular}
  \end{center}
  \caption{Watermark payload.}
  \label{tab:wm_payload}
\end{table}

%% file: service.tex
\section{Manifest Database}
\label{sec:db}
Ideally in the future,  all \Sys manifests and ledger receipts will be delivered as additional metadata with the media objects.
Delivering the receipt along with the media allows the client to quickly validate that the media has been previously authenticated
without contacting the \Sys Service. The widespread use of adding the manifest and receipt to the metadata will most likely require
adoption by one or more multimedia standards bodies.
In the mean time, a client can use the \ManDb to map a media object or chunk to a suitable manifest and receipt.

The \Sys \ManDb contains manifests and receipts.
It is exposed as a public service that lets clients obtain one or more \Sys manifests and receipts that authenticate a published or transcoded media object.
To perform this function efficiently, the \ManDb uses the following indexes: (a) the MediaID delivered via the metadata or a watermark,
and (b) the media ObjectHash or, in the case of streaming media, the cryptographic hashes of all of the contained
chunks (ChunkDigest).

Media players can quickly and easily extract or calculate the ObjectHash or a ChunkDigest from the media, and then use the \ManDb to find a matching manifest and the corresponding receipt.
To validate the legitimacy of any manifest that was retrieved from the \ManDb the following steps will need to occur:
\begin{enumerate}
  \item The contents of the manifest will be hashed by a predetermined cryptographic hash function.
  \item The receipt will then be checked to ensure that it contains the previously calculated hash.
  \item The validator will then validate that the receipt is \textit{endorsed} by  media provenance ledger via a signature over the receipt by the private key of the CCF service.
\end{enumerate}
These steps ensure the validity of the manifest returned by the \ManDb by proving it is produced and endorsed by the media provenance ledger.

The \ManDb can be centralized or distributed.  Because authoritative truth is stored in the ledger, the security requirements for the \ManDb are much less than for the ledger itself.
Note that \Sys manifests do not address problems that arise from more than one publisher signing the same original content – either the same simple object or one or more ChunkDigests.  Similarly, the \Sys Service does not stop a rogue CDN from claiming that one media object is a faithful transformation of an original when in fact it has been maliciously authored.  We believe that these issues can be addressed by a combination of client policies (e.g., only consider the oldest manifest of a media object) and server-side terms-of-service.

\noindent\textbf{Transformation Services.}
A Transformation Service takes one or more media objects and creates a derived object.  A CDN is a simple example: CDNs can take a single media object and re-encodes it into several derived objects with different compression parameters to optimize for bandwidth and network losses.  \Sys manifests support transformation services by allowing entities to indicate the ManifestID of one or more source objects that were used to create the derived object.

Note that a transformation manifest does not in itself guarantee that a derived object is indeed a high-fidelity transformation of a source object. It is entirely possible that the ``purportedly derived'' object is unrelated to the stated original. Trust assessments should involve the entity that signed the transformation manifest.  In the simple case, this might be the original publisher. For example, a media publisher creates a master media object and a dozen copies with different compression factors.
A more complex example might be a CDN acting on behalf of the media publisher.

Policies can be developed for transitive trust that work for common scenarios.  These policies can be enforced with a combination of client- and server-side rules, as well as server-side terms-of-service.
Other entities might create and sign transformation manifests.  For example, a third-party service might use heuristics to compare the semantic content of two videos and create and sign transformation manifests for the videos
that they determine are semantically identical.  Once more, \Sys makes no trust assumptions: it is up to clients to use trust policies that are appropriate for a given scenario.
In the case of streaming manifests, there is no requirement that source-chunks map 1:1 to transformed chunks: chunks are ``natural'' for each stream.

\noindent\textbf{Manifest Revocation.}
As noted previously, CCF's ledger is immutable; once a manifest is stored on the CCF ledger, it cannot be removed. Therefore when
a publisher wants to revoke a manifest from the ledger, it must insert a revocation object to the ledger. To enable efficient queries, the \ManDb deletes this manifest in this case.

%% file: uxSupp.tex
\section{User Experience}
\label{uxSupp}
User experience is a critical part of the \Sys system. \Sys provides three separate
types components to facilitate a good user experience including a modified browser,
browser extensions and example web sites.
\subsection{Modified Chromium Browser}
For media with the embedded manifest, we first created a modified Edge Chromium
browser which included a modified video element. This modification was done to evaluate
the embedded manifest included in a modified video stream.

\subsection{Provenance Browser Extensions}
\Sys includes separate browser extensions for displaying media
using detached manifests and embedded manifests. The goal of the detached manifest
is to allow for the authentication of the media without any modifications to the
standardized media and browser as well as the media transport infrastructure. To
do so, we have created a browser extension that works on both the Chrome and latest
Edge Chromium browsers.

\begin{figure}[tb]
\centering
  \includegraphics[trim = 0.0in 0.0in 0.0in 0.0in,clip,width=1.0\columnwidth]{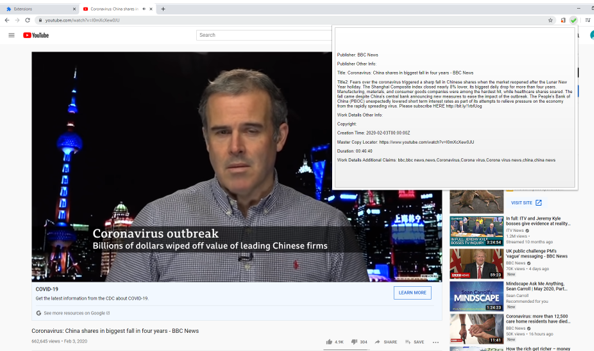}
  \caption{Browser extension which supports a detached manifest. The green check mark in the
  upper right indicates that the video has been authenticated. Clicking on the browser
  extension icon causes the core manifest information to be displayed in the popup window.}
  \label{fig:extension}
\end{figure}

To support media authentication, the \Sys detached manifest browser extension
monitors the web traffic using the webRequest API for a particular site.
Typically, a particular media player embedded in either a browser or an application
streams the media for playback or rendering using HTTP partial-GETs. This process
fetches data from the media server based on different protocols which may vary
depending upon network quality conditions. Since \Sys's detached manifest
requires hashes to be computed on fixed chunk sizes so that the hashes of
the received media matches those in the manifest, \Sys's browser extension
must stream of second copy of the data.
For demonstration purposes, we have implemented the browser extension to
authenticate any video on YouTube after a manifest for that video
has been uploaded to the \SysServ. An example of this browser extension is shown in
the Figure~\ref{fig:extension} for the case of a video which has been registered in
the manifest. For this authenticated video, the browser extension displays a green
check mark indicating that the video's manifest has indeed been located in the
Manifest Database. If the video cannot be authenticated, the browser extension icon
is blank.
If a video can be authenticated, then clicking on the browser extension icon does display
the core manifest information associated with its publication.

The browser extension
for the embedded manifest has a similar user interface. However, the detection mechanism is
much simpler than for the detached manifest because the browser signal is generated by the
new video element in the modified Edge Chromium browser.

\subsection{Example Web Sites}
\begin{figure}[tb]
\centering
  \includegraphics[trim = 0.0in 0.0in 0.0in 0.0in,clip,width=1.0\columnwidth]{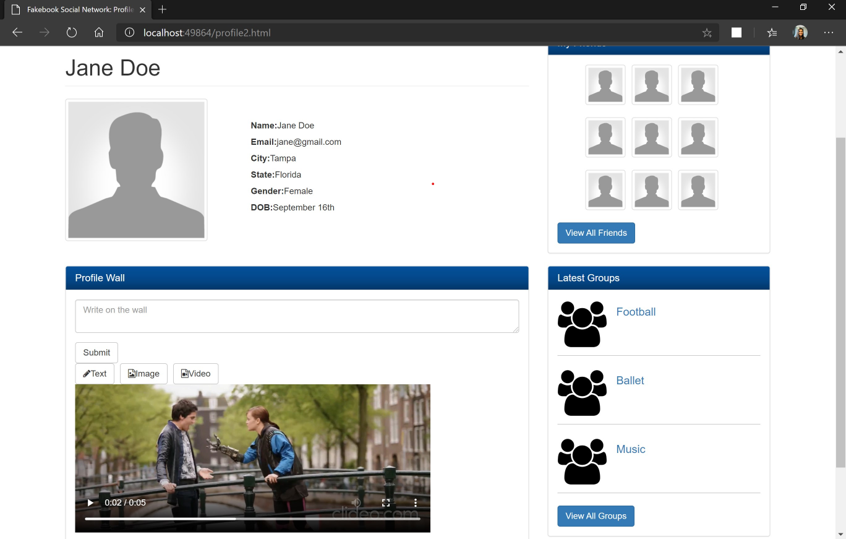}
  \caption{Example of a synthetic social media that can be used to display fine-grained provenance signals.}
  \label{fig:socialmedia}
\end{figure}

\begin{figure}[tb]
\centering
  \includegraphics[trim = 0.0in 0.0in 0.0in 0.0in,clip,width=1.0\columnwidth]{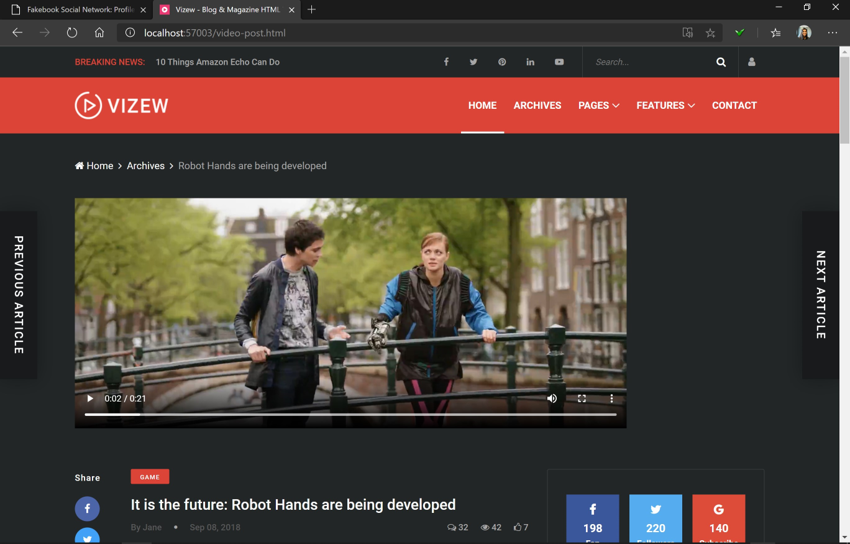}
  %\caption{Demonstration news web site for user study research on provenance signals.}
  \caption{Demonstration news web site for displaying \Sys's provenance signals.}
    \label{fig:news}
\end{figure}

While browser extensions are able to convey whether a single video which is being played
has been authenticated, it has difficulties conveying to a user if two or more video
can authenticated while being played simultaneously. Another challenge is
that it may be difficult for the
user to understand which video on a web page containing multiple videos is being
authenticated. In this case, it may be useful for the web page developer to embed
the provenance signal directly into the web page. To this end, we have developed
two demonstration web sites to display \Sys's authentication information in
a fine-grained setting. The first is a synthetic social media web site shown in
Figure~\ref{fig:socialmedia}, and the second is an example news site that is depicted in Figure~\ref{fig:news}.

%% file: tools.tex
\section{Tools and Governance}
\label{sec:tools}
There are two parts to the authoring and management back-end.
The first part supports the publishing flow.
We have developed tools that create a signed manifest (\Sys Manifest Creation Tool, \Sys Signing Tool),
watermark the media (\Sys Watermark Tool) and record the manifest on a ledger (\Sys Ledger Insertion Tool).
The \Sys Client Provenance Library can be used by the client application to chunk a video and compute the cryptographic
hashes of these chunks.
These tools and tool-chains can be used by an ISP, CDN, or another media editing tool to support ``authenticated transformations'' of an original work, as well as tools that allow authentication information to be added to legacy media (e.g., videos already hosted on YouTube).

The second part of the authoring back-end relates to governance.
We use the Microsoft CCF (Confidential Consortium Framework) technology to maintain a ledger of published works and provide a governance model over it.
CCF provides a flexible governance model, allowing for a group of members to vote on everything from adding and removing users to updating the CCF service code.
If \Sys is adopted to provide media provenance, we will collaborate with our media partners at that time to create a governance model. When additional partners join the partnership, we will use CCF to evolve the governance rules as required. 

%% file: flow.tex
\section{Example Media Publishing Flow}
\label{sec:flow}

\begin{figure}[tb]
\centering
  \includegraphics[trim = 1.0in 6.15in 1.0in 1.0in,clip,width=1.0\columnwidth]{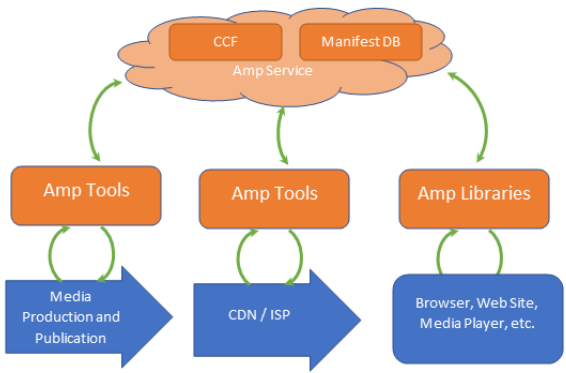}
  \caption{Integration of \Sys tools and services into a media production and distribution pipeline.}
  \label{fig:system}
\end{figure}

The purpose and operation of the various \Sys components is demonstrated by tracing a typical flow of media through the system. A typical flow is depicted in Figure~\ref{fig:system}. The media publishing
flow consists of two phases: publishing and playback. We present below how various \Sys Service components can
be used during the publishing and playback phases.

\subsection*{Publishing}
Assume a content producer generates two media objects: picture.jpg, and video.mp4. The publisher:
\begin{enumerate}
  \item Uses ffmpeg to convert video.mp4 into a set of re-compressions, video[n].mp4, at various quality levels (e.g., using DASH).
  \item Generates a collection of unique ObjectIDs for the objects to be authenticated.
  \item Uses the \Sys Watermarking Tool to insert encoded versions of the ObjectIDs, their PublisherID and the WatermarkPayloadSignature into the watermark payload of the picture.jpg and all videos that are to be published.
  \item Uses the \Sys Manifest Creation Tool to create a manifest for the media objects.
  \item Uses the \Sys Signing Tool to signing the manifest with its publisher's key.
  \item Registers the manifest's cryptographic hash and copyright string with the Media Provenance Ledger using the \Sys Ledger Insertion Tool.
  \item Uploads the manifests to the \Sys \ManDb using the \Sys Ledger Insertion Tool.
  \item Broadcasts (i.e., stages on a web site, etc.) picture.jpg and video[n].mp4.
\end{enumerate}
Optional step for CDNs, ISPs, etc:
\begin{enumerate}
  \item	CDNs take video[n].mp4 and picture.jpg and create further derived copies using steps 4 through 8 except that these manifests refer back to the original \Sys manifest.
\end{enumerate}
\subsection*{Playback}
A client application (e.g., browser, media player, etc.):
\begin{enumerate}
  \item	 Links to the \Sys Client Provenance Library or implements the functionality itself to cryptographically hash a video object's ``chunks'' or simple media object (e.g., JPEG, text).
  \item Consults the \Sys Service to obtain a suitable manifest or manifests.
  \item	Verifies the publisher's signature and the receipt generated by CCF to ensure that the manifest is valid. Successful verification ensures the authenticity of the media.
  \item Displays the authentication information (simple or more complex information) if the media is authenticated.
  \item	Searches for a watermark in the media If a valid manifest is not found in the \ManDb. Next, attempts to validate the media object based on the PublishedID and ObjectID if the WatermarkPayloadSignature is valid.
\end{enumerate}

%% file: eval.tex
\section{Performance Evaluation}
\label{sec:eval}
\noindent\textbf{Media Provenance Ledger.}
We begin by measuring the time required to insert a manifest's relevant data into the \textit{Media Provenance Ledger}.
%This measurement also includes the time needed for generating the receipt for the manifest insertion from CCF.
In this test, we insert strings, which consist of an example 256-bit cryptographic hash of a manifest (ManifestID) and a copyright string, into the ledger.
These data structures do not need to be addressable in CCF since the fact that they are recorded in the ledger is sufficient.
To this end, we measure the maximum sustainable rate at which a manifest's data can be submitted. % and receipts can be obtained.

\emph{Application.}
We built a C++ application that customizes the CCF framework to produce a \textit{Media Provenance Ledger}.
The ledger application is small and can be expressed in several hundred lines of C++ code.
The following is an example of the data that the ledger application stores:
% \begin{verbnobox}[\tiny]
\begin{verbnobox}
{"method": "LOG\_record",
 "params'': {"id": 0, "msg":
   "88c3ba2b25cef698d9ca6775b7fd5c5e
    8bbc246098a55ad51b8078834c4add44
    Copyright (c) CompanyName Corporation.
    All rights reserved."}
}
\end{verbnobox}

\emph{Experimental Setup.}
We ran the performance application in three cluster configurations:
\begin{enumerate}
  \item \emph{Single Azure Region} - Each computer is an Intel(R) Xeon(R) E-2176G CPU @ 3.70GHz, and the application runs inside a 4 core virtual machine.
    \label{lst:single-region}
  \item \emph{2 Geographically distributed Azure Regions} - Each computer is an Intel(R) Xeon(R) E-2176G CPU @ 3.70GHz, and the application runs inside a 4 core virtual machine.
    The computers are every distributed between the east USA and west Europe Azure regions.
    \label{lst:multi-region}
  \item \emph{Emerging hardware} - A cluster that is running in our own datacenter. All computers are under the same 40G switch, and the computers is an Intel(R) Xeon(R) E-2288G CPU @ 3.70GHz which has 8 cores.
    \label{lst:local}
\end{enumerate}
All of these VMs are running Ubuntu 18.04, and the results are shown in Table~\ref{tbl:perf:all}.
We expect that there will be up to 1 billion entries added to the ledger every day, this results in an expected load of 11,575 operations per second.
We can conclude from these results that our implementation of the \textit{Media Provenance Ledger} can comfortably handle this load.
Even with just a few nodes, we can achieve latencies that are low enough to not interfere with the user's experience in media consumption.

\begin{table}[tb]
  \begin{center}
    \begin{tabular}{l@{\hspace{-10pt}}rrrr}
      \toprule
      \textbf{Configuration \ref{lst:single-region}} & & & Throughput (tx/s) & Avg. latency (ms) \\
      \midrule
      \textbf{nodes} & & \\
      % \midrule
      1 & & & 34,316 & 105 \\
      3 & & & 31,828 & 154 \\
      5 & & & 30,763 & 159 \\
      7 & & & 30,013 & 164 \\
      \toprule
      \textbf{Configuration \ref{lst:multi-region}} & & & Throughput (tx/s) & Avg. latency (ms) \\
      \midrule
      \textbf{nodes} & & \\
      1 & & & 34,316 & 105 \\
      3 & & & 32,415 & 244 \\
      5 & & & 31,617 & 245 \\
      7 & & & 30,500 & 248 \\
      % \midrule
      \toprule
      \textbf{Configuration \ref{lst:local}} & & & Throughput (tx/s) & Avg. latency (ms) \\
      \midrule
      \textbf{nodes} & & \\
      1 & & & 57,433 & 80 \\
      3 & & & 52,798 & 131 \\
      5 & & & 52,308 & 132 \\
      7 & & & 49,237 & 140 \\
      \bottomrule
    \end{tabular}
  \end{center}
  \caption{Media Provenance Ledger throughput and latency.}
  \label{tbl:perf:all}
\end{table}

\noindent\textbf{\Sys Service.}
Next, we estimate the maximum scale requirements for the \Sys Service assuming the following
parameters:
\begin{itemize}
\item 10,000 publishers
\item Every publisher uploads 100, 10-minute original video clips uploaded each day
\item The video is divided into 10 second chunks (10 mins is 60 chunks) and each chunk is cryptographically hashed
\item Each original video is transformed into 99 (100-1) variants by the CDN
\end{itemize}

Using these parameters, this translates into
\begin{itemize}
\item 370 million original videos/year
\item 37 billion original and transformed videos/year
\item 22 billion original chunks/year
\item 2.2 trillion total chunks/year
\end{itemize}

Since the \Sys Service is independent of the CCF nodes, we can use large-scale VMs for implementing the index.
If the index is a 32-byte cryptographic hash and 32 bytes of other data (manifest Copyright field), the total index size for all known chunks is 1.4 TBytes. Azure offers VMs with enough memory and disk to hold the index in a single instance, and therefore the index will not require sharding.

If the \Sys Service exceeds these estimates, we can shard the index.
Scaling through sharding is easy: the indices are cryptographic hashes so they will be uniformly distributed.
Therefore, we believe that it will be practical to have the \ManDb indexed on chunk-hashes.

\noindent\textbf{Audio Watermarking.}
\Sys's audio watermarking module inserts a watermark
into the frequency domain coefficients of the audio signal.
It is important to measure the distortion introduced
by the watermark, as we want it to be imperceptible.
Table~\ref{tbl:audio_wm} measures the
Objective Difference Grade (ODG)~\cite{arnold2002subjective} for the audio channel of four different YouTube videos.
The ODG ranges from 0 (no distortion) to -4 (high perceptual distortion).
The mean and standard deviation are computed for five different
trials with 1000 random bits of information inserted using 512
chips per information bit. Preliminary experiments show that watermarking generates
no audible distortions.
\begin{table}[tb]
  \begin{center}
    \begin{tabular}{lll}
      \toprule
      \textbf{YouTube ID} & ODG Mean & ODG Std \\
      \midrule
      XFmn9kmZAWU & -0.74 & 0.0066 \\
      xn\_8UQ1W6\_c & -1.51 & 0.0043 \\
      bF\_nULoyi9o & -0.99 & 0.0039 \\
      iuX826AGXWU & -1.26 & 0.0026 \\
      \bottomrule
    \end{tabular}
  \end{center}
  \caption{Objective difference grade scores for audio watermarking of four different YouTube videos.}
  \label{tbl:audio_wm}
\end{table}

%% file: consortium.tex
\section{\Sys Partnership and Standards}
\label{sec:consortium}

We believe that the proposed \Sys media provenance certification and verification system
can only be successful if it becomes a widely adopted industry standard. Thus, we are
forming a partnership with media organizations and additional technology providers.
We plan to put this collaboration on a formal footing through the formation of an
industry alliance similar to the Alliance for Open Media.  Other companies can join,
either as active contributors or supporters. Such a model can move quickly for
ratification of a more detailed design, with the goal of developing reference code
and performing sufficient testing to assess the efficiency and performance of the
proposed provenance certification and verification system.

A key goal of such an effort should be to promote the development of an open standard,
to motivate fast adoption. We believe that the implementation of a reliable provenance
certification and verification system can be a significant step in increasing trust in
media. It will also benefit the business models of all bona fide entities involved in
the creation and distribution of media.

%% file: discussions.tex
\section{Discussions}
\label{sec:discussions}

It will take a number of years before manifests for a large percentage of online media are stored in \Sys.
We believe this content gap and inability to report on the authenticity of media will be biggest issue with adoption.
We expect that this can be solved in the user-interface such that users are only informed when there is valuable information to provide to them.
At the point when most media that is consumed does have authentication it would become prudent to report that authentication for some media is missing.
Another direction for future work is understanding how provenance information
should best be conveyed to the user as another heuristic for evaluating content
credibility~\cite{zannettou2019web,schwarz2011augmenting}.

One area that \Sys does not address is the detection of fake media.
We believe that the quality of fake media will rapidly improve and become more widely encountered.
Additional fake media detection algorithms will need to be incorporated into the media processing pipeline in the near future.
A number of academic and industry efforts are currently underway to improve the detection of deep fakes.
We see this work as orthogonal to the provenance solution proposed by \Sys, and these detection methods can also be included in the future as part of the \Sys service.

We have designed \Sys to authenticate that a media item was published by a known source.
\Sys is not a digital rights management (DRM) system that is designed to enforce copyright of the media content providers.
Media provenance and \Sys are about verifying the producing entity, not verifying/tracking/authorizing the consuming entity.
While it is possible to use \Sys in this way, functionality such as self-verifiable receipts would work against this, and this is a property we do not intend to change.

%% file: related.tex
\section{Related Work}
\label{sec:related}
Previous related research to the \Sys system and effort span three main areas:
previously proposed provenance systems, other provenance partnerships, and deep fake detection and content generation.

\noindent\textbf{Provenance Systems}.
Provenance systems for the prevention of deep fakes is a new and relatively understudied area.
The provenance-based system that is most closely related to \Sys was recently proposed by Hasan~\cite{Hasan19}.
Like \Sys, this system also employs blockchain. However, it is based on the Ethereum blockchain and smart contracts.
Since \Sys utilizes CCF, it is much more efficient, allowing the speedup of manifest insertion
and queries by several orders of magnitude which is required for widespread deployment.

In addition to~\cite{Hasan19}, several startups have proposed provenance-based systems including: Amber and Witness.
Amber's technology~\cite{amber,Newman19} is aimed at camera manufacturers and adds a cryptographic hash to the video at a user specified rate.
Similar to \Sys, these hashes are then stored on an Ethereum blockchain.

Similarly, Truepic~\cite{Truepic} also provides a photo and verification service where the cryptographic signature is
written to a blockchain.

\noindent\textbf{Provenance Partnerships}. Several other partnerships have been created to ensure the provenance of media.
The New York Times Company is working with IBM on The News Provenance Project~\cite{NewsProv}. This collaboration is also using a blockchain
to provide a provenance solution for media.

The Content Authenticity Initiative is a second partnership with Adobe, The New York Times Company  and Twitter~\cite{CAI}.

Witness is a non-governmental organization which
aims to help ensure that human rights abuses can be documented in a verifiable manner. Witness published the
ProofMode Android application~\cite{ProofMode} in 2017 which stores metadata about images and
videos taken by those seeking to provide evidence of human rights abuses.
The app includes a hash of the media and its metadata along with a cryptographic signature that
helps to ensure the chain of custody.

\noindent\textbf{Deep Fake Detection}. Deep fake detection is an alternate method to provenance solutions and rely
on the algorithmic detection of synthetically generated media. A number of deep fake detection algorithms have been
proposed in the literature.
%Some of the early work in this area include the following works.

In~\cite{Li18_Oculi}, Li et al. describe their realization that
deep fake videos which had been created prior the paper's publication in 2018 often had eyes which failed to blink, which
is natural for humans. Thus, they created an eye blink detector and used it as a proxy to detect deep fake videos.

McCloskey and Albright~\cite{McCloskey18} noted that generative adversarial networks (GANs) fail to accurately reproduce colors that are
captured naturally by photosensitive cells in a camera's sensor. Their approach to detecting deep fakes
is to train a convolutional neural network (CNN) to detect this mismatch in the color.

Face warping artifacts can be introduced during the generation of deep fake videos. Li and Lyu
trained a CNN to detect these artifacts to detect some types of deep fake attacks in~\cite{Li19_FaceWarping}. Similarly,
Yang et al.~\cite{Yang19} also trained a CNN to detect inconsistencies in head poses.

In~\cite{Korshunov18}, Korshunov and Marcel explore trying to jointly use the audio and video, but their experiments indicated that adding
the audio did not help.

In the FaceForensics++ system proposed by R\"ossler et al.~\cite{roessler2019faceforensics++}, the Xception computer vision object recognition model which also employs CNNs were also used to various types of deep fakes. A leader board of deep fake detection algorithms on the FaceForensics++ dataset can be found at~\cite{FFLeader}.

\noindent\textbf{Content Generation}.
Generative adversarial networks were originally proposed by Goodfellow, et al.~\cite{Goodfellow14}.
Several important works~\cite{Karras18,Brock19} have investigated using GANs for large-scale, synthetic image generation.
Recent research in GANs has enabled talking head models to be quickly adapted with just a few frames~\cite{Zakarov19}.

Popular face swap algorithms include Deepfakes~\cite{Deepfakes} and FaceSwap~\cite{FaceSwap}. Facial expressions can be transferred from one person to a person in a video in real-time using the Face2Face algorithm~\cite{thies2016face}.

%% file: conc.tex
\section{Conclusion}
\label{sec:conc}

Synthesized and fake media has become a threat to individuals and private and public
institutions. The threat has increased because of rapid advances in methods for
synthesizing media, coupled with the wide reach of the Web. Fake media has the
potential to significantly undermine trust in media and journalism, threatening the
foundations of democracy. We believe that algorithms for fake media detection will
have limited success in the long term, so we propose the use of provenance
certification and authentication, as that is a fundamental step in increasing trust.

We have proposed, designed, and built a prototype of the \Sys system. \Sys allows
trusted content providers to form one or more consortiums that allow applications
such as a media player or a browser to provide an indication to users that the
source of the content they are viewing has been verified. Beyond the core security
pipeline, human factors and design will play an important role on the success of
\Sys. Inspired by the TLS lock icon, we propose that applications such as browsers
and  media players include UI elements to alert users that the received content can
be traced back to its original source.

For a provenance solution such as \Sys to be successful, it must be formally adopted
by a recognized standards body. We are seeking the development of such standards for
the \Sys system or a variant that provides similar functionality. We also believe
that it is important to open source the code for a widely used provenance system.
Thus, we plan to open source the \Sys system in near future to facilitate its
widespread adoption.

%% file: introSupp.tex
\section{Introduction}
The purpose of this appendix is to provide additional details about the \Sys system design which are not covered in the
submitted paper. In particular, Section~\ref{manifestDetails} includes an extended description of the different types of manifests,
and Section~\ref{sec:man_details} describes \Sys's structures in detail.
Finally, CCF is used to provide
additional functionality which is described in~\ref{CcfSupp}.

%% file: manifestSupp.tex
\section{Manifest Details}
\label{manifestDetails}
\begin{figure}[tb]
\centering
  \includegraphics[trim = 1.0in 4.5in 1.0in 1.0in,clip,width=1.0\columnwidth]{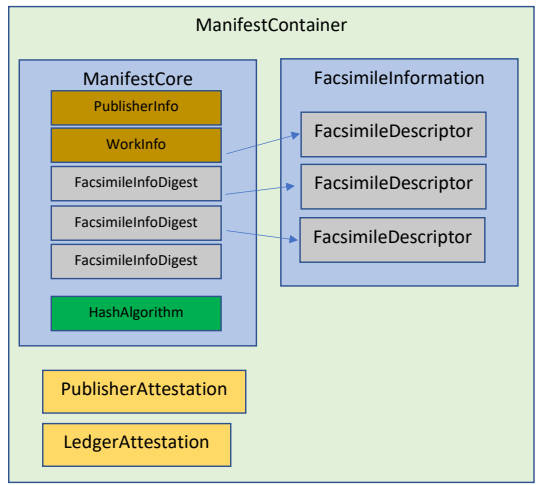}
  \caption{A simplified illustration of a ManifestCore and related data structures. Many fields and some data structures are
omitted for clarity. %The actual structure definitions appear later in this specification.
}
  \label{fig:container}
\end{figure}

This section provides additional details about the full manifest which is depicted in Figure~\ref{fig:container}. A manifest
is actually a container (ManifestContainer) and includes a core manifest called the ManifestCore, a FacssimileInformation to describe
facsimiles of the original media object, and two structures which provide supporting evidence of the publisher (PublisherAttestation) and the
ledger (LedgerAttestation). Before describing the manifest details, we next provide an overview of how to use a manifest.

\subsection{Using Manifests}
Manifests can be created by publishers, re-distributors (CDNs, ISPs), social media platforms, recording
devices, etc. and \Mans are signed by the entity that created them. Manifests can also be countersigned
by cloud services: for example, the CCF cloud service produces a signed receipt to acknowledge
that a Media Manifest has been recorded on a ledger.
Manifests are conventionally JSON or CBOR encoded; the canonicalization rules are described later in
this specification. If the canonicalization rules are followed, then \Mans can be translated back and
forth from JSON to CBOR as needed.
For ease of development, the \Man authoring tools sign both the CBOR representation (using a COSE
signature) and the JSON representation (using a JWT signature). The ManifestID is the hash of the
CBOR-encoded manifest.
Manifests can be delivered to clients as metadata with the actual media objects. However, since the
ecosystem for delivering media is complex, we expect that it will take time for this delivery
infrastructure to be widespread. Considering this, \Sys provides a \ManDb that clients
can use to search for a \Man for a work. There are several ways to query the \ManDb,
including querying by ManifestID, querying by the hash of the entire work, or an array of hashes of
chunks of the work.

\subsection{Authenticating Works}
Each \Man authenticates either precisely one work, or several facsimiles of a work. There are
no technical restrictions on what constitutes a facsimile, but the intention is that facsimiles support the
very common scenario in which web sites, CDNs, etc. prepare a family of media objects (images,
video, audio) that are optimized for different devices and network conditions, but all of which represent
the same content – just not the same exact bits.

Manifests broadly contain two classes of data:

\noindent{\textbf{Metadata:}
This is publisher-assigned data, such as a publisher name and a title for the work.

\noindent{\textbf{Media bindings:}
This describes the facsimiles: for example, cryptographic digests of the media, or subsets/chunks of the
media and media type information.

These fields are described in more detail in the following sections.

\subsection{Metadata}
Most metadata is contained in the structures PublisherInfo and WorkInfo, with the option to include
facsimile-specific information in the FacsimileDescriptor structure.
This specification intentionally limits the metadata that is defined in this structure, and still less is
mandatory. A minimal set of metadata would be the name of the publisher and the name of the work.
If additional metadata needs to be attached, then it can be expressed in the OtherClaims data structures.
The manifest supports an array of OtherClaims structures to be included in PublisherInfo (claims about
the publisher), WorkInfo (claims about the work), FacsimileDescriptor (claims about the facsimile), and
SourceWork (describing the how a source work was transformed to produce a derived work).
OtherClaims allows two sorts of claims to be associated with the manifest. Claim-sets can be embedded
directly into the manifest, or a URI (or other descriptor) can be used to associate claims outside the
manifest. In the case of external claims, OtherData allows the option that the manifest can
cryptographically commit to the external claims by including the hash of the external data in the
OtherClaims structure.
OtherClaims contains a string type descriptor. We expect to define a few standard descriptors such as
``XMP,'' ``EIDR'', ``SCHEMA,'' and then use a DNS-style namespace to allow extensions. For example, the
current manifest tools use the type descriptor ``;com.microsoft.amp.youtube-info'';
to encode YouTube
metadata.

\subsection{Media Bindings}
\subsubsection{Authentication using an Object Digest}
All facsimiles are authenticated by hashing the entirety of the data that constitutes the facsimile: e.g.,
the hash of the entire PDF, JPG, MP4, OGV, etc. file.
Some commonly used multimedia standards allow multiple streams to be packaged in a single object. In
some cases, it still makes sense to authenticate the entire container file or stream. In other cases, a
subset of the underlying media file is authenticated. One important example of this is when the
\Man is packaged in the media file itself – for example, when the \Man is embedded in an
ISO/MPEG container. In all cases, the \Man directly or indirectly specifies exactly what parts of the
media object are hashed.

\subsubsection{Authentication using Chunking}
Most modern media players download and buffer a few seconds of media and then start playing almost
immediately, so authenticating a media object based on the hash of the whole file is inappropriate. To
support progressive/streaming playback of media, the system supports streaming authentication using a
collection of the hashes of ``chunks'' of the media object.
Different media delivery schemes demand different chunking schemes. Two chunking schemes are
currently supported: file-offset-based chunking, and a Merkle-tree based scheme for MP4-containerized
video. Each Facsimile can be authenticated using more than one chunking scheme to allow a single
work to be delivered in multiple ways.

\subsubsection{File-Offset-Based Chunking}
The most common media rendering technology on the web today is the HTML5 \textless video\textgreater\xspace element.
The
simplest way of using an HTML5 video player is to configure the
\textless video\textgreater\xspace
element to fetch video data
from a URL. In this case, the \textless video\textgreater\xspace
element performs a sequence of HTTP partial-GET operations to
fetch the video data.
File-offset-based chunking can be used to do progressive authentication in this case: the manifest
contains an array of hashes of (say) 256KB chunks of the underlying video file, and the video player or
browser calculates video-stream hashes and checks that they match a manifest.
File-offset-based hashing can also work with Adaptive Bitrate (ABR) Streaming in some circumstances.
ABR on the web is enabled by video player logic (often a JavaScript library running in the web page)
fetching audio and video data from a collection of files encoded at different bandwidths. File-offset-based
chunking still works in this case: each of the underlying video files is chunked, hashed, and
encoded in the manifest.
The SimpleChunkList data structure is used to encoded file-offset-based chunking. SimpleChunkLists
contain an array of hashes and the size of the underlying chunks. The size of each chunk is recorded in
the manifest, but we will additionally define some standard lengths to enable chunk-hashes to be
calculated by clients when they do not yet have a valid manifest. The final chunk in a file may be less
than the chunk size.

\subsubsection{MP4-Container Hashing and Merkle Tree Authentication}
The MP4 ISO/IEC container format is a widely used standard for encoding any sort of media object in a
file or stream. MP4 defines ``box'' types for holding multimedia data and metadata. For the purposes of
this discussion, the following box types are important:

\begin{itemize}
  \item MOOV: Basic stream metadata: one per container
  \item MDAT: Video or audio data: typically, a few seconds
  \item MOOF: Describes the samples in the subsequent MDAT
\end{itemize}

The simplest fragmented MP4 container contains \{MOOV [MOOF, MDAT] +\}, but most containers have
additional boxes.
MP4-Container-chunking defines a chunk as a subset of the MOOF data that defines the sample,
together with the corresponding video data: i.e., the MDAT.

Chunk hashes defined in this way can be embedded in a ManifestCore using the
MerkleTreeAuthenticator, described next.

\subsubsection{Merkle Tree Authentication}
Typical chunk sizes for fragmented MP4 are a few seconds long, so the chunk hash data can be quite
large. If the authentication data is encoded as a simple array, then the array of chunk hashes must be
available in its entirety before authentication can begin. The MerkleTreeAuthenticator is an alternative
representation of the chunk-hashes that allows authentication to begin when only a subset of the
authentication data is available. This is achieved by encoding part of the authentication data in the
manifest, and additional ``evidence'' in the media stream itself. Together, these allow a player to check
that a media chunk is consistent with a manifest.
This form of authentication is supported by encoding the authentication data as a Merkle hash tree. A
Merkle Tree, depicted in Figure~\ref{fig:merkle}, is a binary tree of hashes, where the leaves of the tree are the digests of the [MOOF MDAT]
samples, and each row in the tree is the hash of the data or hashes in the row beneath.

\begin{figure*}[tb]
\centering
  \includegraphics[trim = 1.0in 6.15in 1.0in 1.0in,clip,width=1.2\columnwidth]{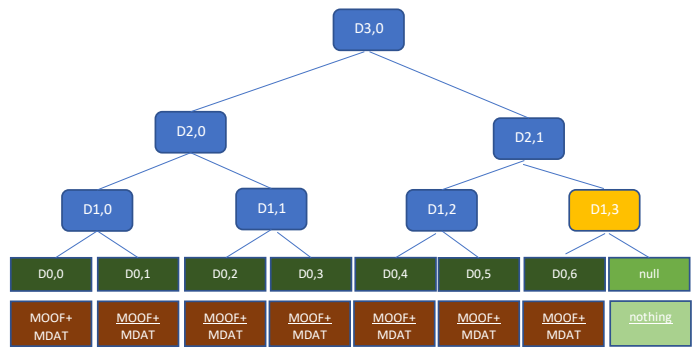}
  \caption{Merkle Hash Tree formed over multimedia data. The ``leaves'' of the hash tree are the hashes of the media samples,
and each row in the hash tree is formed from the hash of the concatenation of the two hashes in the lower layer. The top-hash
is called the root of the hash tree. If the number of samples is not a power of two, the leaves of the ``missing'' samples at the
end of the file are null, and are processed according to the rules in this section.}
  \label{fig:merkle}
\end{figure*}

The Merkle Tree authenticator is encoded in two parts, which are typically distributed separately. The
actual Media Manifest contains one row of hashes from the tree: for example, the D2,0 and D2,1 digests in
the figure above. This would be sufficient to the authenticate the video data as long as the player can
read and hash all of the data leading up to D2,0 or D2,1, but (in this example) the player would have to
read, chunk, and hash half of the file before authentication could begin.
To avoid the need for excessive read-ahead, the media can be distributed with the relevant missing
parts of the tree, so that the player can validate that a particular chunk hash is consistent with the
manifest. For example, in the figure above, to prove that the first sample is consistent with D 2,0 the
evidence would be D0,1 and D1,2 because these hash values can be used to form the missing parts of the
tree.

The tree is formed as follows.
The depth of the tree is determined by the number of chunks in the file. In general, the number of leaf
hashes is not a power of two.  In such cases, the tree depth is calculated by rounding up the number of
leaf hashes to the next power of two.  For example, if there are 5 chunks, then this rounds up to 8,
which leads to a tree depth of 4, including the leaves of the tree.
The general rules for forming the tree (in both the power-of-two, and non-power of two case) are as
follows:
\begin{enumerate}
  \item The leaf hashes are formed from the hash of the chunk data.
  \item The ``hash'' of non-present chunk is termed null.
\end{enumerate}

To form intermediate node hashes in the tree:
\begin{enumerate}
  \item If both inputs are non-null, then output = Hash (LHS|RHS)
  \item If one input (RHS) is null, then output = LHS
  \item If both inputs are null, then output = null 
\end{enumerate}
    
The MerkleTreeAuthenticator data structure encodes one row of the hash tree in the Media Manifest,
omitting null values. Encoding of the evidence hashes is described in the next section.

\noindent{\textbf{Encoding Evidence in an MP4 Container.}
The evidence that allows a player to determine that a chunk is consistent with an associated manifest is
encoded in an ISO/MP4 UUID-box called a ChunkIntegrityBox.
%
%\noindent{\textbf{Chunk Integrity Box.} The Chunk Integrity Box
The ChunkIntegrityBox enables verification of a set of samples when combined with the manifest.
%Definition

\begin{tabbing}
Box \= Type: `uuid' \\
\% (Big-Endian Bytes) \\
Box Extended Type: 469d22dfe1924defa71ef4c9f2ce3e71 \\
Container: Movie Fragment Box (`traf') \\
Mandatory: No \\
Quantity: Zero or one \\
Syntax \\
class ChunkIntegrityBox extends \\
FullBox(`uuid', 469d22dfe1924defa71ef4c9f2ce3e71, \\
version=0, flags=0) \\
\{\\
\>   unsigned int(8) \= hash\_tree\_id; \\
\>   unsigned int(16)\= hash\_location; \\
\>   unsigned int(8) \= hash\_size; \\
\>   unsigned int(8) \= hash\_count; \\
\>   \{\\
\>     unsigned int(8)[hash\_size] hash; \\
\>   \}[hash\_count] \\
\} \\
\end{tabbing}
\noindent The ChunkIntegrityBox fields are as follows:
\begin{itemize}
  \item hash\_tree\_id: The index into the list of hashed streams in the manifest
  \item hash\_location: The zero-based chunk index
  \item hash\_size: The size, in bytes, of the hash value
  \item hash[] Every non-null hash from the tree required to get from this chunk’s node to one of the
nodes found in the \Man.  These hashes are sequenced from leaf to root.
\end{itemize}

\subsection{Adaptive Bitrate Streaming}
The MPEG-DASH and Microsoft Smooth Streaming are adaptive bitrate (ABR) streaming formats that
allow a client player to select between different encodings of the same video object. Stream selection
can happen when playing starts but, if network conditions change, can also happen during playback.
Under the covers, these streaming standards are usually enabled by creating a set of underlying
compressed media files and dynamically assembling them into HLS or DASH objects with CMAF (MP4)
chunks. The individual files are encoded using different bandwidths/compression ratios, and, for each
bandwidth, the original video is usually split into shorter files to allow client players to switch
bandwidths every few seconds by fetching from a different source file.
Adaptive streaming is supported by a set of ManifestCore strucures by treating each of the separately encoded
constituent files as a Facsimile. In some cases, this might be a Transformation Manifest with a
back-pointer to the manifest for an original high-definition file that was used to create the ABR streams, and
in other cases the ABR streams will all be authenticated using a simple (non-transformation) manifest.
%An example of a \Man for an adaptive stream set is included at the end of this document.

\subsection{Transformation Manifests}
Transformation Manifests are used to authenticate Works that are transformed from other Works.
Transformation Manifests can be authored by the same publisher that created the original work, can be
authored by an entity operating on behalf of another entity (e.g. a CDN), or can be created by a
completely unrelated entity, tool, or person.\footnote{Trust assessments when several parties are involved are not discussed here.} Such manifests allow an entity to apply a transformation
to a work, establish the original work as its source, and make a signed claim this transformation does not
alter the meaning of the content of the original Work. The manifest does not itself prove this assertion
automatically but provides an auditable trail through which the assertion could be challenged. How such
a challenge would be resolved is beyond the scope of this work; the manifest only ensures the
transforming entity is accountable for transformed Works it releases.

Transformation Manifests differ from original work manifests in that they specify the ManifestID of the
source work or works used to create the derived work, and also include the nature of the
transformation applied. The primary initial scenario enabled by Transformation Manifests is
re-encoding of a media object after the original manifest is created. However, we have allowed for future
extensibility to express more complex sorts of derivation such as editing and media object composition.
Such an extension of Transformation Manifests may allow for the meaning of the original work to be
altered, but in a specific and documented way they assert is acceptable and does not alter the meaning
of the transformed content. For example, a derivative work in the form of a news report might use a clip
of a newsworthy event, and the producing entity could both assert the originality of its own content and
make a claim that the clip of the event being described is unaltered, or itself transformed in some
acceptable way, such as transcoded, or decorated with the entity's chyron or watermark.

\subsection{Distributing Manifests and Manifest Containers}
A simplified representation of a ManifestCore and related data structures in illustrated in Figure~\ref{fig:container}.
The central data structure that cryptographically authenticates media is called the ManifestCore. A
ManifestCore directly contains some data items, and cryptographic commitment to external data
structures that may be distributed with the manifest or by other means.
The ManifestCore uses commitments/hashes rather than embedding the data structure directly when the
supplemental data is not always required. For example:

\begin{itemize}
  \item The facsimile media authentication information is encoded in one or more external
FacsimileDescriptors. This allows a media object to be distributed with only the
FacsimileDescriptors that are relevant. For example, if a video object is encoded in WEBM and
MP4, and each is encoded in 5 different bit rates and resolutions, this is 10 facsimiles. If a player
is just playing one of these streams, then only the appropriate FacsimileDescriptor needs to be
available to authenticate the stream.
  \item There are a wide range of media metadata formats, and there is a wide range of data that a
publisher might want to associate with a work; some of which the publisher might not want to
distribute. Commitment can be used to attach supplemental metadata to a \Man.
\end{itemize}

The consequence of this is that a ManifestCore always needs additional data structures before it can
be used to authenticate media. The ManifestContainer data structure is an envelope that allows a
ManifestCore to be distributed with supplemental data structures that allow a work to be
authenticated. Note that the ManifestCore cannot be modified after it is created because the MediaID
would change and signatures would break. However, ManifestContainers (each of which contains a
ManifestCore) can be freely created with just the data needed for the intended purpose.
ManifestContainers can also contain signature blocks and certificates from the publisher
(PublisherAttestation and LedgerAttestation).

\subsection{Signing Manifests}
Manifests are typically signed by the originator (publisher, redistributor, social media platform, etc.) and
may be counter-signed by distributed ledger services. Manifest signatures are performed over the hash
of a canonical representation of the manifest.
%At the time of writing,
JSON and CBOR representations
are used by different parts of the system, so the manifest is signed twice: once to produce a JWT
signature block (JSON) and once to produce a COSE signature block (CBOR). A PublisherAttestation
optionally allows the signer certificate or certificate chain to be bundled in the ManifestContainer.

\subsection{Canonicalization}
ManifestIDs and signatures are over JSON or CBOR canonical encodings. JSON canonicalization follows
the IEFT JCS draft. CBOR canonicalization follows RFC7049. COSE signatures follow RFC8152. 

%% file: mandet.tex
\section{Manifest Structures}
\label{sec:man_details}
This section includes the detailed definitions for some of the key \Sys manifest structures. 
%The entire manifest specification is 43 pages and beyond the scope of this supplementary material.
%
\subsection{ManifestContainer}
A ManifestContainer (Table~\ref{tab:ManifestContainer}) is a holder for information needed to authenticate a media object.

The ManifestContainer structure contains all the information necessary to validate a work. ManifestCore is the central structure: it is usually hashed and signed, in which case publisher signing information is held in PublisherAttestation. If the manifest is registered on a public ledger/blockchain then additional evidence from the service provider can be stored in LedgerAttestation.

The actual multimedia data is hashed to allow it to be authenticated during playback or forensic analysis. The media hashes are not stored directly in the ManifestCore, but instead are stored in FacsimileDescriptor structures inside FacsimileInfo, with one FacsimileDescriptor structure per facsimile. The FacsimileDescriptors are cryptographically bound to the manifest by hashing. Keeping the FacsimileDescriptors separate allows ManifestContainers to be smaller in the case where a manifest is expected to be used with just one or a few facsimiles.

\begin{table*}
    \begin{center}
      \begin{tabular}{|l|l|l|}
        \hline
        Name & Type & Description \\
        \hline
        \hline
        Version	number & The structure version. & This document describes version 1. \\
        CoreManifest &	ManifestCore & ManifestCore authenticates a media object and associated metadata. To authenticate \\
        & & a media object, a ManifestCore and one or more FacsimileDescriptors (embedded in  \\
        & &  the FacsimileInformation) data structure are required. \\
        FacsimileInfo & FacsimileInformation & A container for one or more FacsimileDescriptors that cryptographically authenticate  \\
        & & media objects. Note that FacsimileInformation may contain descriptors for a subset \\
        & & of the Facsimiles described by the manifest (to reduce storage and bandwidth when \\
        & &  not all FacsimileDescriptors are required.) \\
        PublisherAttestation & PublisherAttestation & Manifest signatures and certificates from the publisher (optional) \\
        LedgerAttestation &	LedgerAttestation & Manifest signatures and certificates from ledger (or other) services (optional) \\
        ManifestLocator	 & string & An optional string that helps locate the manifest or additional \\
        & & FacsimileDescriptors (optional) \\
        \hline
      \end{tabular}
  \end{center}
  \caption{ManifestContainer structure description.}
  \label{tab:ManifestContainer}
\end{table*}

\subsection{ManifestCore}
A ManifestCore (Table~\ref{tab:ManifestCore}) cryptographically authenticates a single work or a set of facsimiles of a work (e.g., a set of JPG images with different sizes and compression ratios, or, for video, different bandwidth encodings, different video frame sizes, and different encoding schemes). Supported work/media types include video, audio, image, text, PDF, HTML.

A ManifestCore structure will often be packaged inside an enveloping ManifestContainer structure. The enclosing ManifestContainer contains additional information to validate a media object, as well as signatures from the publisher and other parties.

The ManifestCore does not directly contain the FacsimileDescriptors that authenticate a facsimile; instead, a ManifestCore contains (essentially) an array of hashes of FacsimileDescriptors. The FacsimileDescriptors themselves are stored outside the ManifestCore - often in an enclosing FacsimileInformation structure. This saves storage and bandwidth if a manifest is being used to authenticate just one or a subset of the defined Facsimiles.

In addition to cryptographically authenticating a work, a ManifestCore contains an optional publisher assigned metadata identifying the publisher (PublisherInfo), and the work being authenticated (WorkInfo). These structures can also reference external metadata.

The ManifestCore allows the expression of ``authorized derivation'' of a work by services such as social platforms, CDNs, or publishing tools. To support this, a ManifestCore can contain a back-pointer to other ManifestCores called Origin Manifests. If the work is a simple transcoding of another work, then this will point to the manifest for the original work. If the work is a composite of several originals, then the ManifestCore can point back to several originals.

All cryptographic digests in a ManifestCore and related structures must use the hashing algorithm described in ManifestCore (HashingAlgorithm).
The ManifestID is the hash of a canonical representation of a manifest. Currently, this is the hash of a canonical CBOR-encoding of the manifest.

\begin{table*}
    \begin{center}
      \begin{tabular}{|l|l|l|}
        \hline
        Name & Type & Description \\
        \hline
        \hline
        Version	number & The structure version. & This document describes version 1. \\
        SerialNumber &	byte[] &	Statistically unique / random serial number for the manifest \\
        DigestAlgorithm & string & All hashes in this manifest and the contained data structures use the \\
        & & algorithm stated here \\
        MediaID & byte[] & A publisher-assigned quasi-unique ID for the work or family of works. The MediaID \\
        & & can attached to works (e.g., in a metadata field in the file) or encoded \\
        & & using a watermark. \\
        CreationTime & date-time & The date/time when this manifest was created. Note that WorkInfo can specify a \\
        & & different time for the creation of the work. \\
        Publisher & PublisherInfo & Information about the publisher (or redistributor) that created this manifest \\
        Work & WorkInfo & Information about the work or works described by this manifest \\
        FacsimileInfoDigests & byte[][] & An array of hashes of FacsimileInfo structures that are typically delivered in an \\
        & & enveloping ManifestContainer. The FacsimileInfo structures authenticate the media \\
        & & objects described by this manifest. \\
        OriginManifests & SourceWorkInfo & If the manifest is a derived work (transcoding or composite edited work) this data \\
        & & structure contains the original manifest of manifests, as well as the transformations \\
        & & that were applied. (optional) \\
        \hline
      \end{tabular}
  \end{center}
  \caption{ManifestCore structure description.}
  \label{tab:ManifestCore}
\end{table*}

In this structure, ``FacsimileInfoDigests'': {``type'': [``array'',``null''],``items'': {``type'': [``string'',``null'']}}.

\subsection{PublisherInfo}
The PublisherInfo structure in Table~\ref{tab:PublisherInfo} is a container for information about the publisher or redistributor of this manifest.

\begin{table*}
    \begin{center}
      \begin{tabular}{|l|l|l|}
        \hline
        Name & Type & Description \\
        \hline
        \hline
        Name & string & The name of the publisher \\
        OtherInfo & string & Any other information that the publisher needs to associate with the work or works (optional) \\
        AdditionalClaims & OtherClaims[] & Any other information about the publisher that should be associated with this manifest (optional) \\
        \hline
      \end{tabular}
  \end{center}
  \caption{PublisherInfo structure description.}
  \label{tab:PublisherInfo}
\end{table*}

\subsection{OtherClaims}
OtherClaims (Table~\ref{tab:OtherClaims}) is a container for additional claims to be associated with a publisher, work, facsimile, or transformation.
ManifestCores natively support a minimal amount of metadata. Publishers may choose to include or reference additional metadata about the work, the facsimile, the transformation, or the publisher using this data structure.
Two types of extension are supported: (1) EmbeddedClaims is any string-encoded data that is embedded in manifest itself, and (2) ExternalClaims is a pointer (e.g. an URL, file name or GUID) to an external data object. Optionally, ExternalClaimsDigest can contain the digest of external data if its integrity must be protected.

\begin{table*}
    \begin{center}
      \begin{tabular}{|l|l|l|}
        \hline
        Name & Type & Description \\
        \hline
        \hline
        Name & string & The name of the publisher \\
        ClaimSort & string	& Publisher chosen identifier for the sort of metadata encoded in this record.  \\
        % E.g. ``com.contoso.stuff'' ``XMP'', or ``EIDR'' \\
        EmbeddedClaims &	string & String encoding of additional metadata (optional) \\
        ExternalClaims	& string & A locator (URI, etc.) of external metadata (optional) \\
        ExternalClaimsDigest & byte[] & Optional digest of the external metadata. This can be used if the additional metadata is stable \\
        & & and the publisher wishes to cryptographically commit to the exact metadata at the time of \\ 
        & &  manifest creation. If this is not required, then this field should be omitted or null. (optional) \\
        \hline
      \end{tabular}
  \end{center}
  \caption{OtherClaims structure description.}
  \label{tab:OtherClaims}
\end{table*}

Specifically, ``EmbeddedClaims'': {``type'': [``string'', ``null'']} and  ``ExternalClaimsDigest'': {``type'': [``string'', ``null'']}.

\subsection{WorkInfo}
The WorkInfo structure (Table~\ref{tab:WorkInfo}) is a container for publisher- or redistributor-provided information about a work. This information is the same for all facsimiles described by a ManifestCore.
\begin{table*}
    \begin{center}
      \begin{tabular}{|l|l|l|}
        \hline
        Name & Type & Description \\
        \hline
        \hline
        Title & string & The name or title of the work or family of works \\
        Title2 & string	& Additional name/title information (optional) \\
        OtherInfo & string & Optional publisher-chosen data (optional) \\
        Copyright & string & A copyright notice for the work or family of works (optional) \\
        CreationTime & date-time & Publisher-chosen original publication or creation time. This need not be the same as the \\
        & & manifest creation time (optional) \\
        MasterCopyLocator & string & A stable URI, etc. of a master original (facsimiles may have their own Facsimile \\
        & & locators) (optional) \\
        Duration & number & If the work is video or audio this can be the length of the work in 100ns (1e-7 secs) \\
        & & units (optional) \\
        AdditionalClaims & OtherClaims[] & Other publisher-provided metadata. (optional) \\
        \hline
      \end{tabular}
  \end{center}
  \caption{WorkInfo structure description.}
  \label{tab:WorkInfo}
\end{table*}

\subsection{SourceWorkInfo}
SourceWorkInfo (Table~\ref{tab:SourceWorkInfo}) identifies one or more source works that were used to produce a derived work.

\begin{table*}
    \begin{center}
      \begin{tabular}{|l|l|l|}
        \hline
        Name & Type & Description \\
        \hline
        \hline
        OriginManifests & SourceWork[] & An array of identifiers for the source works, \\
        & & and how the source works were processed to create the derived work. \\
        \hline
      \end{tabular}
  \end{center}
  \caption{SourceWorkInfo structure description.}
  \label{tab:SourceWorkInfo}
\end{table*}

\subsection{SourceWork}
SourceWork (Table~\ref{tab:SourceWork}) identifies the source work in a transformation manifest. It also describes the type of transformation performed and may optionally contain details about the exact transformation applied.

\begin{table*}
    \begin{center}
      \begin{tabular}{|l|l|l|}
        \hline
        Name & Type & Description \\
        \hline
        \hline
        OriginManifest & ManifestReference & A reference to the manifest of the origin work \\
        DerivationType & DerivationSort & Describes the transformation of the source work to form \\
        & & the derived work: e.g. a simple transcoding. \\
        AdditionalClaims & OtherClaims[] & Any other information about the transformation that was applied \\
        & & to the original to produce the derived work transformation, \\
        & & for example, EIDR claims. (optional) \\
        \hline
      \end{tabular}
  \end{center}
  \caption{SourceWork structure description.}
  \label{tab:SourceWork}
\end{table*}

\subsection{ManifestReference}
A ManifestReference (Table~\ref{tab:ManifestReference}) is a description of a ManifestCore that is stored elsewhere. A ManifestLocator MUST contain the ManifestID of the referenced manifest and may optionally include the URI of a service by which the Manifest can be obtained.

\begin{table*}
    \begin{center}
      \begin{tabular}{|l|l|l|}
        \hline
        Name & Type & Description \\
        \hline
        \hline
        Version & number & The structure version. This document describes version 1. \\
        ManifestLocator & string & An optional field to encode a service, file, etc. that can \\
        & & be used to locate the referenced manifest (optional) \\
        ManifestID & TypedDigest & The ManifestID (manifest digest) of the referenced manifest \\
        \hline
      \end{tabular}
  \end{center}
  \caption{ManifestReference structure description.}
  \label{tab:ManifestReference}
\end{table*}

\subsection{TypedDigest}
TypedDigest (Table~\ref{tab:TypedDigest}) is a container class for a typed digest. Most digests used in ManifestCore-related data structures are simple byte-arrays with the hash algorithm defined the associated ManifestCore. This data structure is used when typed digests are required.

\begin{table*}
    \begin{center}
      \begin{tabular}{|l|l|l|}
        \hline
        Name & Type & Description \\
        \hline
        \hline
        DigestAlgorithm & string & The digest/hash algorithm used to create this digest \\
        DigestValue & byte[] & The digest/hash value \\
        \hline
      \end{tabular}
  \end{center}
  \caption{TypedDigest structure description.}
  \label{tab:TypedDigest}
\end{table*}

\subsection{DerivationSort}
The DeriviationSort (Table~\ref{tab:DerivationSort}) is an enumeration that describes the type of transformation of the original work used to form a derived work.

\begin{table*}
    \begin{center}
      \begin{tabular}{|l|l|l|}
        \hline
        Name & Type & Description \\
        \hline
        \hline
        Transcoded & 1 & The derived work is a simple transcoding of the original work \\
        CompleteCopy & 2 & The entire original work is included in the derived work \\
        PartialCopy & 3 & Part of the original is included in the derived work \\
        EditedCopy & 4 & One or more named editing operations have been applied to the original to produce the derived work \\
        \hline
      \end{tabular}
  \end{center}
  \caption{DerivationSort structure description.}
  \label{tab:DerivationSort}
\end{table*}

\subsection{FacsimileInformation}
The FacsimileInformation structure (Table~\ref{tab:FacsimileInformation}) is a container structure for one or more TaggedFacsimileDescriptors.

\begin{table*}
    \begin{center}
      \begin{tabular}{|l|l|l|}
        \hline
        Name & Type & Description \\
        \hline
        \hline
        Version	& number & The structure version. This document describes version 1. \\
        Records	& TaggedFacsimileDescriptor[] & An array of FacsimileDescriptors tagged with an \\
        & & index that is the location in ManifestCore.FacsimileInfoDigests. \\
        \hline
      \end{tabular}
  \end{center}
  \caption{FacsimileInformation structure description.}
  \label{tab:FacsimileInformation}
\end{table*}

\subsection{TaggedFacsimileDescriptor}
A TaggedFacsimileDescriptor (Table~\ref{tab:TaggedFacsimileDescriptor}) is a container for a FacsimileDescriptor. The index is the array index of the hash of the FacsimileDescriptor in the associated ManifestCore.

\begin{table*}
    \begin{center}
      \begin{tabular}{|l|l|l|}
        \hline
        Name & Type & Description \\
        \hline
        \hline
        Index & number & The zero-based array index into ManifestCore.FacsimileInfoDigests \\
        & & that contains the digest of this FacsimileDescriptor \\
        Facsimile & FacsimileDescriptor & The crypotographic descriptor of a Facsimile \\
        \hline
      \end{tabular}
  \end{center}
  \caption{TaggedFacsimileDescriptor structure description.}
  \label{tab:TaggedFacsimileDescriptor}
\end{table*}

\subsection{FacsimileDescriptor}
A facsimile is a particular encoding or representation of a work and is represented by a FacsimileDescriptor (Table~\ref{tab:FacsimileDescriptor}). A ManifestCore can describe one facsimile, or a collection of facsimiles that the publisher deems equivalent: e.g., a set of videos with different encoding schemes or parameters.
Non-streaming media is cryptographically bound to the hash/digest of the complete work. Streaming media can also be progressively authenticated. Progressive authentication is supported using an array of digests of ``chunks'' of the media stream.
Different scenarios are best supported by different chunking schemes so the ChunkAuthenticator data structures come in several forms. The simplest is file-offset based chunking. Other chunking schemes are also defined, and more will be added as needed.

\begin{table*}
    \begin{center}
      \begin{tabular}{|l|l|l|}
        \hline
        Name & Type & Description \\
        \hline
        \hline
        FacsimileMajorType & FacsimileType & Media type of this facsimile (video, audio, muxed, etc.) \\
        ContainerType & string & The name of the file/container format for this multimedia, \\
        & & e.g., JPG or MP4. \\
        EncodingInformation & string & String-encoded encoding scheme and parameters for this particular  \\
        & & facsimile. If this is a muxed stream, then this will contain the video  \\
        & & encoder info, and EncodingInformation2 will contain the audio \\
        & & encoding info \\
        EncodingInformation2 & string & If this Facsimile contains more than one media type, then this is the  \\
        & & secondary type. E.g. the audio encoder type for an AV muxed \\
        & & stream. (optional) \\
        Length & number & Length, in bytes, of the facsimile \\
        ObjectDigest & byte[] & Digest of the entire work using the hash algorithm in the containing \\
        & & Amp Manifest \\
        FacsimileLocator & string & Any other information about facsimile that should be associated \\
        & & with this (optional) \\
        ObjectContainers & string & If missing or null, then the data hashes to obtain the ObjectDigest \\
        & & is the entire object - e.g., the JPG or MP4 file. If the data to be hashed \\
        & &  is wrapped in a container format and not all of the data in the  \\
        & & enveloping file/stream should be hashed,  then this field which   \\
        & & containers/streams should be hashed (placeholder/todo) (optional) \\
        AdditionalClaims & OtherClaims[] & Any other data that the publisher wishes to associate \\
        & & with the facsimile. (optional) \\
        ChunkData & Array of any & One or more chunked representations of the facsimile. Only needed for \\ 
        & SimpleChunkListAuthenticator & progressive authentication of streaming media objects, ) \\
        & IsoBoxAuthenticator & otherwise null or omitted. (optional \\
        & MerkleTreeAuthenticator &  \\
        \hline
      \end{tabular}
  \end{center}
  \caption{FacsimileDescriptor structure description.}
  \label{tab:FacsimileDescriptor}
\end{table*}

\subsection{FacsimileType}
A FacsimileType (Table~\ref{tab:FacsimileType}) is the type of media object of a Facsimile. Note that a video stream may be decomposed into separate video, audio, and muxed facsimiles. 

\begin{table*}
    \begin{center}
      \begin{tabular}{|l|l|l|}
        \hline
        Name & Type & Description \\
        \hline
        \hline
        Unknown	& 0	& Facsimile type is not known or is not specified \\
        MuxedAV	& 1	& Multiplexed AV stream \\
        Video	& 2	& Video stream (no audio) \\
        Audio	& 3	& Audio stream \\
        Image	& 4	& Any sort of image \\
        Text	& 5	& Any sort of text \\
        \hline
      \end{tabular}
  \end{center}
  \caption{FacsimileType structure description.}
  \label{tab:FacsimileType}
\end{table*}

\subsection{ChunkAuthenticator}
ChunkAuthenticator (Table~\ref{tab:ChunkAuthenticator}) is the base class for various ways that the chunks of a streaming work can be authenticated. Chunks are always authenticated by the hash of a chunk, but the definition of a chunk (e.g., its size, or how chunk boundaries are established) can vary. Concrete variations are defined by different derived structures with different ChunkingScheme values.

\begin{table*}
    \begin{center}
      \begin{tabular}{|l|l|l|}
        \hline
        Name & Type & Description \\
        \hline
        \hline
        ChunkingScheme & number & This tag indicates the actual type of this structure \\
        NumChunks & number & The number of chunks described by this authenticator \\
        ChunkDigest & byte[][] & An ordered array of chunk-hashes starting from the beginning \\
        & & of the work. All ChunkAuthenticators have a list of chunk digests, \\
        & & but specific authenticators may have additional data that describe exactly \\
        & & what each chunk maps to (e.g. file offset-based, I-frame-to-I-frame, etc.) \\
        \hline
      \end{tabular}
  \end{center}
  \caption{ChunkAuthenticator structure description.}
  \label{tab:ChunkAuthenticator}
\end{table*}

\subsection{SimpleChunkListAuthenticator} 
SimpleChunkListAuthenticator  (Table~\ref{tab:SimpleChunkListAuthenticator}) describes file/stream-offset based chunking. For example, if ChunkSize in 1MiByte, then the first chunk is the first MiByte of the media object/file, the second chunk is the second MiByte, etc. The last chunk in a file/stream can be smaller.

\begin{table*}
    \begin{center}
      \begin{tabular}{|l|l|l|}
        \hline
        Name & Type & Description \\
        \hline
        \hline
        ChunkingScheme & number & SimpleChunkList is ChunkingScheme 1 \\
        ChunkSize & number & All chunks are this size (optional) \\
        NumChunks & number & The number of chunks described by this authenticator) \\
        ChunkDigest & byte[][] & An ordered array of chunk-hashes starting from the beginning of the work. All \\
         & & ChunkAuthenticators have a list of chunk digests, but specific authenticators may have additional  \\
         & & data that describe exactly what each chunk maps to (e.g. file offset-based, I-frame-to-I-frame, etc.) \\
        \hline
      \end{tabular}
  \end{center}
  \caption{SimpleChunkListAuthenticator structure description.}
  \label{tab:SimpleChunkListAuthenticator}
\end{table*}

\subsection{IsoBoxAuthenticator}
IsoBoxAuthenticator (Table~\ref{tab:IsoBoxAuthenticator}) describes chunks of multimedia data encoded in an MPEG/ISO container.

\begin{table*}
    \begin{center}
      \begin{tabular}{|l|l|l|}
        \hline
        Name & Type & Description \\
        \hline
        \hline
        ChunkingScheme & number & Iso-box chunking is ChunkingScheme 2 \\
        NumChunks & number & The number of chunks described by this authenticator \\
        ChunkDigest & byte[][] & An ordered array of chunk-hashes starting from the beginning of the work. All \\
        & & ChunkAuthenticators have a list of chunk digests, but specific authenticators may have additional \\
        & & data that describe exactly what each chunk maps to (e.g. file offset-based, I-frame-to-I-frame, etc.) \\
        \hline
      \end{tabular}
  \end{center}
  \caption{IsoBoxAuthenticator structure description.}
  \label{tab:IsoBoxAuthenticator}
\end{table*}

\subsection{MerkleTreeAuthenticator}
MerkleTreeAuthenticator supports chunk authentication using a Merkle hash tree. This style of chunk authentication minimizes the amount of data that needs to be downloaded to authenticate the first chunk (including the case where playback starts in the middle of the file.)
We use the following terminology: the top-hash is the root of the tree, also referred to as row zero. The two children of the top-hash are row one, and so on. The hashes in the final row in the hash tree are also called the leaves of the tree.
The leaves of the tree are the hashes of the media chunks using the hash algorithm specified in the manifest. If the number of chunks is not a power of two, then the tree is padded with leaves with zero-values: e.g., 32 bytes of zero in the case of a sha256 hash tree.
The number of hashes in the ChunkAuthenticator to support the MerkleTreeAuthenticator must always be a power of two (todo, we could relax this), and the hash algorithm used is the algorithm in the containing manifest.
The hash tree is typically split into two parts at a row called the split-row. The upper part of the tree - from the root to the split-row - is encoded in the MerkleTreeAuthenticator (Table~\ref{tab:MerkleTreeAuthenticator}) and hence distributed with the manifest. The lower part of the tree is distributed with the media itself.
The MerkleTreeAuthenticator contains the row of hashes at the split-row, with the row number given by the EncodedRow field. The number of hashes encoded will always be a power of two. If needed, the tree up to the root can be calculated by repeated hashing. Rows below the split-row can be derived from the media itself, or can be derived from the media and fragments of the hash tree called ``evidence'' sent by other means.
It is out of scope of the manifest specification to describe how evidence is encoded, but we expect that each media chunk will be distributed with an array of hashes that allow clients to verify that the hash of a chunk of media data is one of the leaves of the complete hash tree.

\begin{table*}
    \begin{center}
      \begin{tabular}{|l|l|l|}
        \hline
        Name & Type & Description \\
        \hline
        \hline
        ChunkingScheme & number & Hash-Tree-chunking is ChunkingScheme 3 \\
        EncodedRow & number & The row of the tree that is encoded in this authenticator. Zero means that only the root \\
        & & hash is encoded, 1 means that the pair of hashes at row 1 of the Merkle tree is encoded. -1 means \\
        & & that the hashes are the leaf hashes. \\
        NumChunks & number & The number of chunks described by this authenticator  \\
        ChunkDigest & byte[][] & An ordered array of chunk-hashes starting from the beginning of the work. All \\
        & & ChunkAuthenticators have a list of chunk digests, but specific authenticators may have additional data \\
        & & that describe exactly what each chunk maps to (e.g. file offset-based, I-frame-to-I-frame, etc.) \\
        \hline
      \end{tabular}
  \end{center}
  \caption{MerkleTreeAuthenticator structure description.}
  \label{tab:MerkleTreeAuthenticator}
\end{table*}

\subsection{PublisherAttestation}
PublisherAttestation (Table~\ref{tab:PublisherAttestation}) is a container class for publisher-created signatures, etc.

\begin{table*}
    \begin{center}
      \begin{tabular}{|l|l|l|}
        \hline
        Name & Type & Description \\
        \hline
        \hline
        CoseSignatureToken & byte[] & COSE Signature1 signature block (optional) \\
        JsonWebToken & string & String-encoded JSON signature block (optional) \\
        PemEncodedCertificates & string[] & Certificate chain for the signing key ordered from the \\
        & & self-signed root, through the subordinate CAs, to the key used to sign the manifest (optional) \\
        \hline
      \end{tabular}
  \end{center}
  \caption{PublisherAttestation structure description.}
  \label{tab:PublisherAttestation}
\end{table*}

\subsection{LedgerAttestation}
LedgerAttestation (Table~\ref{tab:LedgerAttestation}) is a container class for ledger-created signatures, etc.

\begin{table*}
    \begin{center}
      \begin{tabular}{|l|l|l|}
        \hline
        Name & Type & Description \\
        \hline
        \hline
        LedgerAttestationValue & string & (optional) \\
        \hline
      \end{tabular}
  \end{center}
  \caption{LedgerAttestation structure description.}
  \label{tab:LedgerAttestation}
\end{table*}

%% file: ccfSupp.tex
\section {CCF Details}
\label{CcfSupp}
Manifests are recorded on a public blockchain using CCF.
CCF operates the public ledger (i.e., blockchain) of published works,
essentially a list of manifests,
relying on a distributed network of replicas running on trusted hardware and synchronized using Practical Byzantine Fault Tolerance (PBFT)~\cite{PBFT}  or Raft~\cite{RAFT}.
CCF supports the registration of new manifests and issues signed manifest receipts.
These receipts complement the producer's signatures; they enable any media consumers to independently verify that the work they receive has been published with the corresponding metadata.
CCF also supports online querying and validation of ledger transactions and their endorsing certificates, as well as the transparent governance of the service by a consortium of media producers.

\paragraph{Governance} CCF provides a flexible governance model.
This allows for \Sys{} to define the governance by writing scripts in scripting languages such as the Lua~\cite{Lua} or JavaScript~\cite{javascript} languages.
These scripts specify rules for actions such as adding new members, adding or removing users, adding and removing nodes from the system, user access control, etc.
The specifics of the governance model will be defined as part of the media consortium that controls \Sys{}, and these rules will evolve with time by modifying the governance scripts.

\paragraph{Trust and Integrity} CCF is designed to support two different types of consensus algorithms including
Crash Fault Tolerance (CFT) and Byzantine Fault Tolerance (BFT). The CFT variant that CCF supports is a
modified version of Raft~\cite{RAFT}, and the variant of BFT implemented by CCF is a modified version of
Practical Byzantine Fault Tolerance (PBFT)~\cite{PBFT}.
Raft leverages trusted execution environments (TEEs) and specifically Intel's SGX. Its trust  model is that a single TEE compromise destroys both confidentiality and integrity.
By using this trust model, CCF is able to utilize a variant of Raft which can handle malicious attacks as long as Intel's SGX is not compromised.
PBFT is a consensus algorithm that can make progress if less than ${}^1{\mskip -5mu/\mskip -3mu}_3$ of the nodes are actively malicious. PBFT's trust model is that a single TEE compromise destroys confidentiality but f+1 compromises (in a 3f+1 network) are required to destroy integrity.
This distinction means that even if some of the CCF nodes, which are running in a SGX enclave, are compromised, the Media Provenance Ledger will not lose integrity.
This added security comes at an increased performance and latency cost when committing data to the ledger.
Critically, both of these consensus protocols offer finality.
This property states that once a transaction has been committed, it cannot be reverted.
Furthermore, a CCF receipt provides an additional finality proof.

\paragraph{Distributed Execution} CCF utilizes TEEs to ensure that the operator of the Media Provenance Ledger is not able to perform malicious acts on the service.
This is designed to provide the \Sys{} consortium members with confidence that they can run the ledger service in a cloud datacenter, and that an operator (such as Azure) cannot compromise the service's confidentiality or integrity.